\documentstyle[10pt,epsf,epsfig,dp_delphititle,lineno]{dp_delphi}
\makeindex
\def\DpPaperGroup{EP}
\def\DpPaperRef{2000-068}
\def\DpDate{25 February 2000}
\def\DpAuthors{DELPHI Collaboration}
\def\DpSubmit{(Phys.Lett. B485(2000)45)}
\def\DpTitle{{Measurement and Interpretation of Fermion-Pair Production \\
 at LEP Energies of \\ 183 and 189~GeV}}
\def\DpComment{ }
\def\DpEMail{ }

\newcommand {\capsty}{\normalsize}
\newcommand {\ssize}{\scriptsize}
\newcommand {\GeV} {\mbox{$\mathrm{GeV}$}}

\newcommand {\pbarn} {\mbox{$\mathrm{pb}$}}

\newcommand {\MZ} {\mbox{$\mathrm{M}_{\mbox{\ssize{Z}}}$}}

\newcommand {\Afbe} {\mbox{$\mathrm{A_{\mbox{\ssize FB}}}^{e}$}}
\newcommand {\Afbm} {\mbox{$\mathrm{A_{\mbox{\ssize FB}}}^{\mu}$}}
\newcommand {\Afbt} {\mbox{$\mathrm{A_{\mbox{\ssize FB}}}^{\tau}$}}

\newcommand {\lplm} {\mbox{$l^{+}l^{-}$}}
\newcommand {\ee} {\mbox{$e^{+}e^{-}$}}

\newcommand {\mumu} {\mbox{$\mu^{+}\mu^{-}$}}
\newcommand {\eeg} {\mbox{$e^{+}e^{-}(\gamma)$}}
\newcommand {\mumug} {\mbox{$\mu^{+}\mu^{-}(\gamma)$}}
\newcommand {\tautau} {\mbox{$\tau^{+}\tau^{-}$}}
\newcommand {\tautaug} {\mbox{$\tau^{+}\tau^{-}(\gamma)$}}

\newcommand {\eeee} {\mbox{$e^+e^-\rightarrow e^+e^-$}}

\newcommand {\eeeeg} {\mbox{$e^+e^-\rightarrow e^+e^-(\gamma)$}}

\newcommand {\eemumu} {\mbox{$e^+e^-\rightarrow \mu^+\mu^-$}}

\newcommand {\eemumug} {\mbox{$e^+e^-\rightarrow \mu^+\mu^-(\gamma)$}}
\newcommand {\eetautau}{\mbox{$e^+e^-\rightarrow \tau^+\tau^-$}}
\newcommand {\eettg} {\mbox{$e^+e^-\rightarrow \tau^+\tau^-(\gamma)$}}
\newcommand {\eeqqg} {\mbox{$e^+e^-\rightarrow q\bar{q}(\gamma)$}}

\newcommand {\eell} {\mbox{$e^+e^-\rightarrow l^+l^-$}}
\newcommand {\eeff} {\mbox{$e^+e^-\rightarrow f\overline{f}$}}
\newcommand {\eeqq} {\mbox{$e^+e^-\rightarrow q\overline{q}$}}
\newcommand {\eemm} {\mbox{$e^+e^-\ra \mu^+\mu^-$}}
\newcommand {\eett} {\mbox{$e^+e^-\ra \tau^+\tau^-$}}

\newcommand {\ZZpair} {\mbox{$\mathrm{ZZ}$}}
\newcommand {\WWpair} {\mbox{$\mathrm{W^{+}W^{-}}$}}
\newcommand {\Zgst} {\mbox{$\mathrm{Z}\gamma^{*}$}}

\newcommand {\epair} {\mbox{$e^{+}e^{-}$}}
\newcommand {\mupair} {\mbox{$\mu^{+}\mu^{-}$}}
\newcommand {\taupair} {\mbox{$\tau^{+}\tau^{-}$}}

\newcommand {\acol} {\theta_{acol}}
\newcommand {\mydeg} {\mbox{$^\circ$}}

\def\see{\sigma_{ee}}
\def\smu{\sigma_{\mu\mu}}
\def\stau{\sigma_{\tau\tau}}

\newcommand{\dsdcth}{\mbox{${d\sigma}/{d\cos\theta}$}}

\newcommand {\sqsps} {{\sqrt{s'}/\sqrt{s}}}
\newcommand{\ra}{\mbox{$\rightarrow$}}

\newcommand{\sqs}{\mbox{$\protect\sqrt{s}$}}
\newcommand{\spr}{\mbox{s$^{\prime}$}}
\newcommand{\sqsp}{\mbox{$\protect\sqrt{s^{\prime}}$}}
\newcommand{\TeV}{\mbox{$\mathrm{TeV}$}}
\newcommand{\upb}{\mbox{$\mathrm{pb}$}}
\newcommand{\susyfy}[1]{\mbox{$\stackrel{\sim}{#1}$}}
\newcommand{\msneut}{\mbox{$m_{\tiny{\susyfy{\nu}}}$}}
\newcommand{\etal}{{\it et al.\/}\ }
\newcommand{\ZFITTER}{\mbox{ZFITTER}}

\newcommand{\TOPAZZERO}{\mbox{TOPAZ0}}

\newcommand{\KORALZ}{\mbox{KORALZ}}

\newcommand {\snul} {\tilde{\nu_{\ell}} }

\newcommand {\Zprime} {\mbox{$\mathrm{Z}^{'}$}}
\newcommand {\thtzzp} {\mbox{$\Theta_{\mathrm{Z} \mathrm{Z}^{'}}$}}
\newcommand {\MZp} {\mbox{$\mathrm{M}_{\mathrm{Z}^{'}}$}}
\newcommand {\thtzzplim} {\mbox{$\Theta_{\mathrm{Z} \mathrm{Z}^{'}}^{limit}$}}
\newcommand {\MZplim} {\mbox{$\mathrm{M}_{\mathrm{Z}^{'}}^{limit}$}}

\begin{document}
\makeatletter
\newcount\@tempcntc
\def\@citex[#1]#2{\if@filesw\immediate\write\@auxout{\string\citation{#2}}\fi
  \@tempcnta\z@\@tempcntb\m@ne\def\@citea{}\@cite{\@for\@citeb:=#2\do
    {\@ifundefined
       {b@\@citeb}{\@citeo\@tempcntb\m@ne\@citea\def\@citea{,}{\bf ?}\@warning
       {Citation `\@citeb' on page \thepage \space undefined}}%
    {\setbox\z@\hbox{\global\@tempcntc0\csname b@\@citeb\endcsname\relax}%
     \ifnum\@tempcntc=\z@ \@citeo\@tempcntb\m@ne
       \@citea\def\@citea{,}\hbox{\csname b@\@citeb\endcsname}%
     \else
      \advance\@tempcntb\@ne
      \ifnum\@tempcntb=\@tempcntc
      \else\advance\@tempcntb\m@ne\@citeo
      \@tempcnta\@tempcntc\@tempcntb\@tempcntc\fi\fi}}\@citeo}{#1}}
\def\@citeo{\ifnum\@tempcnta>\@tempcntb\else\@citea\def\@citea{,}%
  \ifnum\@tempcnta=\@tempcntb\the\@tempcnta\else
   {\advance\@tempcnta\@ne\ifnum\@tempcnta=\@tempcntb \else \def\@citea{--}\fi
    \advance\@tempcnta\m@ne\the\@tempcnta\@citea\the\@tempcntb}\fi\fi}
 
\makeatother
\begin{titlepage}
\pagenumbering{roman}
\CERNpreprint{\DpPaperGroup}{\DpPaperRef} 
\date{{\small\DpDate}} 
\title{\DpTitle} 
\address{\DpAuthors} 
\begin{shortabs} 
\noindent
%
\noindent
An analysis of the data collected in 1997 and 1998 with the DELPHI 
detector at $e^+e^-$ collision energies close to 183 and 189 
GeV was performed in order to extract the hadronic and leptonic fermion--pair
cross--sections, as well as the leptonic forward--backward asymmetries and 
angular distributions.
The data are used to put limit on contact interactions between fermions, 
the exchange of R-parity violating SUSY sneutrinos, 
$\Zprime$ bosons and the existence of gravity in extra dimensions.
\end{shortabs}
\vfill
\begin{center}
\DpSubmit \ \\ 
\DpComment \ \\
\DpEMail \ \\
\end{center}
\vfill
\clearpage
\headsep 10.0pt
\addtolength{\textheight}{10mm}
\addtolength{\footskip}{-5mm}
\begingroup
%
\newcommand{\DpName}[2]{\hbox{#1$^{\ref{#2}}$},\hfill}
\newcommand{\DpNameTwo}[3]{\hbox{#1$^{\ref{#2},\ref{#3}}$},\hfill}
\newcommand{\DpNameThree}[4]{\hbox{#1$^{\ref{#2},\ref{#3},\ref{#4}}$},\hfill}
\newskip\Bigfill \Bigfill = 0pt plus 1000fill
\newcommand{\DpNameLast}[2]{\hbox{#1$^{\ref{#2}}$}\hspace{\Bigfill}}
%
\footnotesize
\noindent
\DpName{P.Abreu}{LIP}
\DpName{W.Adam}{VIENNA}
\DpName{T.Adye}{RAL}
\DpName{P.Adzic}{DEMOKRITOS}
\DpName{Z.Albrecht}{KARLSRUHE}
\DpName{T.Alderweireld}{AIM}
\DpName{G.D.Alekseev}{JINR}
\DpName{R.Alemany}{VALENCIA}
\DpName{T.Allmendinger}{KARLSRUHE}
\DpName{P.P.Allport}{LIVERPOOL}
% ...
\normalsize
\endgroup
\titlefoot{Department of Physics and Astronomy, Iowa State
     University, Ames IA 50011-3160, USA
    \label{AMES}}
\titlefoot{Physics Department, Univ. Instelling Antwerpen,
     Universiteitsplein 1, B-2610 Antwerpen, Belgium \\
     \indent~~and IIHE, ULB-VUB,
     Pleinlaan 2, B-1050 Brussels, Belgium \\
     \indent~~and Facult\'e des Sciences,
     Univ. de l'Etat Mons, Av. Maistriau 19, B-7000 Mons, Belgium
    \label{AIM}}
\titlefoot{Physics Laboratory, University of Athens, Solonos Str.
     104, GR-10680 Athens, Greece
    \label{ATHENS}}
\titlefoot{Department of Physics, University of Bergen,
     All\'egaten 55, NO-5007 Bergen, Norway
    \label{BERGEN}}
\titlefoot{Dipartimento di Fisica, Universit\`a di Bologna and INFN,
     Via Irnerio 46, IT-40126 Bologna, Italy
    \label{BOLOGNA}}
\titlefoot{Centro Brasileiro de Pesquisas F\'{\i}sicas, rua Xavier Sigaud 150,
     BR-22290 Rio de Janeiro, Brazil \\
     \indent~~and Depto. de F\'{\i}sica, Pont. Univ. Cat\'olica,
     C.P. 38071 BR-22453 Rio de Janeiro, Brazil \\
     \indent~~and Inst. de F\'{\i}sica, Univ. Estadual do Rio de Janeiro,
     rua S\~{a}o Francisco Xavier 524, Rio de Janeiro, Brazil
    \label{BRASIL}}
\titlefoot{Comenius University, Faculty of Mathematics and Physics,
     Mlynska Dolina, SK-84215 Bratislava, Slovakia
    \label{BRATISLAVA}}
\titlefoot{Coll\`ege de France, Lab. de Physique Corpusculaire, IN2P3-CNRS,
     FR-75231 Paris Cedex 05, France
    \label{CDF}}
\titlefoot{CERN, CH-1211 Geneva 23, Switzerland
    \label{CERN}}
\titlefoot{Institut de Recherches Subatomiques, IN2P3 - CNRS/ULP - BP20,
     FR-67037 Strasbourg Cedex, France
    \label{CRN}}
\titlefoot{Now at DESY-Zeuthen, Platanenallee 6, D-15735 Zeuthen, Germany
    \label{DESY}}
\titlefoot{Institute of Nuclear Physics, N.C.S.R. Demokritos,
     P.O. Box 60228, GR-15310 Athens, Greece
    \label{DEMOKRITOS}}
\titlefoot{FZU, Inst. of Phys. of the C.A.S. High Energy Physics Division,
     Na Slovance 2, CZ-180 40, Praha 8, Czech Republic
    \label{FZU}}
\titlefoot{Dipartimento di Fisica, Universit\`a di Genova and INFN,
     Via Dodecaneso 33, IT-16146 Genova, Italy
    \label{GENOVA}}
\titlefoot{Institut des Sciences Nucl\'eaires, IN2P3-CNRS, Universit\'e
     de Grenoble 1, FR-38026 Grenoble Cedex, France
    \label{GRENOBLE}}
\titlefoot{Helsinki Institute of Physics, HIP,
     P.O. Box 9, FI-00014 Helsinki, Finland
    \label{HELSINKI}}
\titlefoot{Joint Institute for Nuclear Research, Dubna, Head Post
     Office, P.O. Box 79, RU-101 000 Moscow, Russian Federation
    \label{JINR}}
\titlefoot{Institut f\"ur Experimentelle Kernphysik,
     Universit\"at Karlsruhe, Postfach 6980, DE-76128 Karlsruhe,
     Germany
    \label{KARLSRUHE}}
\titlefoot{Institute of Nuclear Physics and University of Mining and Metalurgy,
     Ul. Kawiory 26a, PL-30055 Krakow, Poland
    \label{KRAKOW}}
\titlefoot{Universit\'e de Paris-Sud, Lab. de l'Acc\'el\'erateur
     Lin\'eaire, IN2P3-CNRS, B\^{a}t. 200, FR-91405 Orsay Cedex, France
    \label{LAL}}
\titlefoot{School of Physics and Chemistry, University of Lancaster,
     Lancaster LA1 4YB, UK
    \label{LANCASTER}}
\titlefoot{LIP, IST, FCUL - Av. Elias Garcia, 14-$1^{o}$,
     PT-1000 Lisboa Codex, Portugal
    \label{LIP}}
\titlefoot{Department of Physics, University of Liverpool, P.O.
     Box 147, Liverpool L69 3BX, UK
    \label{LIVERPOOL}}
\titlefoot{LPNHE, IN2P3-CNRS, Univ.~Paris VI et VII, Tour 33 (RdC),
     4 place Jussieu, FR-75252 Paris Cedex 05, France
    \label{LPNHE}}
\titlefoot{Department of Physics, University of Lund,
     S\"olvegatan 14, SE-223 63 Lund, Sweden
    \label{LUND}}
\titlefoot{Universit\'e Claude Bernard de Lyon, IPNL, IN2P3-CNRS,
     FR-69622 Villeurbanne Cedex, France
    \label{LYON}}
\titlefoot{Univ. d'Aix - Marseille II - CPP, IN2P3-CNRS,
     FR-13288 Marseille Cedex 09, France
    \label{MARSEILLE}}
\titlefoot{Dipartimento di Fisica, Universit\`a di Milano and INFN-MILANO,
     Via Celoria 16, IT-20133 Milan, Italy
    \label{MILANO}}
\titlefoot{Dipartimento di Fisica, Univ. di Milano-Bicocca and
     INFN-MILANO, Piazza delle Scienze 2, IT-20126 Milan, Italy
    \label{MILANO2}}
\titlefoot{Niels Bohr Institute, Blegdamsvej 17,
     DK-2100 Copenhagen {\O}, Denmark
    \label{NBI}}
\titlefoot{IPNP of MFF, Charles Univ., Areal MFF,
     V Holesovickach 2, CZ-180 00, Praha 8, Czech Republic
    \label{NC}}
\titlefoot{NIKHEF, Postbus 41882, NL-1009 DB
     Amsterdam, The Netherlands
    \label{NIKHEF}}
\titlefoot{National Technical University, Physics Department,
     Zografou Campus, GR-15773 Athens, Greece
    \label{NTU-ATHENS}}
\titlefoot{Physics Department, University of Oslo, Blindern,
     NO-1000 Oslo 3, Norway
    \label{OSLO}}
\titlefoot{Dpto. Fisica, Univ. Oviedo, Avda. Calvo Sotelo
     s/n, ES-33007 Oviedo, Spain
    \label{OVIEDO}}
\titlefoot{Department of Physics, University of Oxford,
     Keble Road, Oxford OX1 3RH, UK
    \label{OXFORD}}
\titlefoot{Dipartimento di Fisica, Universit\`a di Padova and
     INFN, Via Marzolo 8, IT-35131 Padua, Italy
    \label{PADOVA}}
\titlefoot{Rutherford Appleton Laboratory, Chilton, Didcot
     OX11 OQX, UK
    \label{RAL}}
\titlefoot{Dipartimento di Fisica, Universit\`a di Roma II and
     INFN, Tor Vergata, IT-00173 Rome, Italy
    \label{ROMA2}}
\titlefoot{Dipartimento di Fisica, Universit\`a di Roma III and
     INFN, Via della Vasca Navale 84, IT-00146 Rome, Italy
    \label{ROMA3}}
\titlefoot{DAPNIA/Service de Physique des Particules,
     CEA-Saclay, FR-91191 Gif-sur-Yvette Cedex, France
    \label{SACLAY}}
\titlefoot{Instituto de Fisica de Cantabria (CSIC-UC), Avda.
     los Castros s/n, ES-39006 Santander, Spain
    \label{SANTANDER}}
\titlefoot{Dipartimento di Fisica, Universit\`a degli Studi di Roma
     La Sapienza, Piazzale Aldo Moro 2, IT-00185 Rome, Italy
    \label{SAPIENZA}}
\titlefoot{Inst. for High Energy Physics, Serpukov
     P.O. Box 35, Protvino, (Moscow Region), Russian Federation
    \label{SERPUKHOV}}
\titlefoot{J. Stefan Institute, Jamova 39, SI-1000 Ljubljana, Slovenia
     and Laboratory for Astroparticle Physics,\\
     \indent~~Nova Gorica Polytechnic, Kostanjeviska 16a, SI-5000 Nova Gorica, Slovenia, \\
     \indent~~and Department of Physics, University of Ljubljana,
     SI-1000 Ljubljana, Slovenia
    \label{SLOVENIJA}}
\titlefoot{Fysikum, Stockholm University,
     Box 6730, SE-113 85 Stockholm, Sweden
    \label{STOCKHOLM}}
\titlefoot{Dipartimento di Fisica Sperimentale, Universit\`a di
     Torino and INFN, Via P. Giuria 1, IT-10125 Turin, Italy
    \label{TORINO}}
\titlefoot{Dipartimento di Fisica, Universit\`a di Trieste and
     INFN, Via A. Valerio 2, IT-34127 Trieste, Italy \\
     \indent~~and Istituto di Fisica, Universit\`a di Udine,
     IT-33100 Udine, Italy
    \label{TU}}
\titlefoot{Univ. Federal do Rio de Janeiro, C.P. 68528
     Cidade Univ., Ilha do Fund\~ao
     BR-21945-970 Rio de Janeiro, Brazil
    \label{UFRJ}}
\titlefoot{Department of Radiation Sciences, University of
     Uppsala, P.O. Box 535, SE-751 21 Uppsala, Sweden
    \label{UPPSALA}}
\titlefoot{IFIC, Valencia-CSIC, and D.F.A.M.N., U. de Valencia,
     Avda. Dr. Moliner 50, ES-46100 Burjassot (Valencia), Spain
    \label{VALENCIA}}
\titlefoot{Institut f\"ur Hochenergiephysik, \"Osterr. Akad.
     d. Wissensch., Nikolsdorfergasse 18, AT-1050 Vienna, Austria
    \label{VIENNA}}
\titlefoot{Inst. Nuclear Studies and University of Warsaw, Ul.
     Hoza 69, PL-00681 Warsaw, Poland
    \label{WARSZAWA}}
\titlefoot{Fachbereich Physik, University of Wuppertal, Postfach
     100 127, DE-42097 Wuppertal, Germany
    \label{WUPPERTAL}}
\addtolength{\textheight}{-10mm}
\addtolength{\footskip}{5mm}
\clearpage
\headsep 30.0pt
\end{titlepage}
%
\pagenumbering{arabic} 
\setcounter{footnote}{0} %
\large

\section{Introduction}

Results are presented from the analyses of fermion--pair 
final states collected in 1997 and 1998 by the
DELPHI experiment~\cite{ref:delphidet} at centre--of--mass energies, $\sqs$,
close to 183 and 189~$\GeV$. 
Measurements of cross--sections for inclusive hadronic, 
electron--positron pairs, 
muon--pair and tau--pair final states are given, together with
leptonic forward--backward asymmetries. These results 
complement those obtained from data collected in 1995 and 1996 
at lower collision energies from 130 to $172~\GeV$ \cite{ref:delphi_130_172}.
Polar angle distributions of $\mumu$ and $\tautau$ events recorded at
$\sqs \sim 183$ and 189~$\GeV$ are also given.

The measurements of the cross--sections and forward--backward asymmetries 
together with the results presented in \cite{ref:delphi_130_172} and 
from LEP running in the vicinity of the 
\mbox{Z--resonance}~\cite{ref:delphi91,ref:delphi95}, are used to update the 
searches for new physics involving contact interactions, R-parity violating 
SUSY, and additional neutral gauge bosons given in~\cite{ref:delphi_130_172}. 
In addition, the measurements presented in this paper are used to search 
for possible effects of gravity proposed in theories with large extra 
dimensions.

Results on fermion--pair production at LEP at collision energies from 
130 to $189 \: \GeV$ from the other LEP experiments together with 
limits derived from these results, can be found in \cite{ref:otherlep}.

The measurements of cross--sections, forward--backward asymmetries and
angular distributions are given in section~\ref{sec:measurements}.
The interpretations of the data are presented in 
section~\ref{sec:interpretation}. 
A summary and conclusions are given in section~\ref{sec:summary}.
Further details concerning the data analysis, including the event selection 
and the theoretical and technical details of the searches for new physics 
can be found in~\cite{ref:delphi_130_172}.

\section{Measurements of cross--sections and asymmetries}
\label{sec:measurements}

\subsection{Luminosity and centre--of--mass energy determination}

The luminosity analysis of the data collected during LEP 
operation in 1997 and 1998 followed closely the one described 
in~\cite{ref:delphi_130_172}. The total experimental systematic 
uncertainty on the integrated luminosity determination amounts
to $0.50\%$, to be combined with a $0.25\%$ uncertainty reflecting 
the precision of the theoretical calculations underlying the 
computation of the cross--section visible in the luminometers. 
The luminosities for the analysis of the inclusive hadronic final states
were 52.80 and 155.21 $\upb^{-1}$ for $\sqs \sim 183$ and 189 \GeV\ 
respectively.
Estimates of the mean centre--of--mass energies led
to values of (182.65 $\pm$ 0.05) and (188.63 $\pm$ 0.05) 
$\GeV$~\cite{ref:lepenergy}. There are small differences in the 
luminosities and mean centre--of--mass energies for the other channels 
due to the selection of different running periods for analysis, based on the
performance of the subdetectors of DELPHI.

\subsection{Kinematical definition of signal}

Cross--sections and forward--backward asymmetry
measurements are given for different ranges of the reduced centre--of-mass 
energy, $\sqsp$: for 
hadronic final states an {\it{inclusive}} sample, $\sqsps > 0.10$, and a 
{\it{non--radiative}} sample, $\sqsps > 0.85$; for muon and
tau final states an {\it {inclusive}} sample, $\sqsp > 75 \: \GeV$, and a 
{\it {non--radiative}} sample, $\sqsps > 0.85$. For electron--positron
final states, a cut on the acollinearity\footnote{The acollinearity angle 
between two particles is defined as 
$\cos\acol = - \boldmath{p_{1}.p_{2}}/|\boldmath{p_{1}}||\boldmath{p_{2}}|$
where $\boldmath{p_{1}}$ and $\boldmath{p_{2}}$ are the 3--momenta of the
particles.} angle between the electron and positron,
$\acol < 20\mydeg$, was applied, corresponding approximately to a cut of 
$\sqsps > 0.85$.

The methods of estimating $\sqsp$ correspond to slightly different definitions
of this variable. For \mumu\ and  the \tautau\ final states, 
$\sqsp$ is the invariant mass of the muons or tau-leptons in the final state. 
For the inclusive hadronic final states, the estimated \sqsp\ can be treated 
in theoretical predictions to be the invariant mass of the $s$-channel 
propagator.

For the \ee\ final state the measured cross--sections and forward--backward
asymmetries are for the electron and positron both within the acceptance
$44\mydeg < \theta < 136\mydeg$. 
For the \mumu\ and \tautau\ final states, the cross-sections and asymmetries 
were extrapolated to $4\pi$ acceptance using samples of events generated with 
KORALZ \cite{ref:koralz}. In the calculations of KORALZ there is no 
interference between initial state and final state radiation. Corrections
to the extrapolation for this interference were determined using the 
semi-analytical calculations of ZFITTER \cite{ref:zfitter}, in which the 
interference was computed to $\mathcal{O}(\alpha)$, and applied to the
results. To account for missing higher
order corrections, a systematic uncertainty of half the correction was taken.
For the inclusive hadronic states, where the events are selected over the 
full solid angle, any correction for the interference 
between initial and final state radiation was estimated to be negligibly
small compared to the precision of the measurement.

\subsection{Improvements to analyses}

The analyses of cross--sections for $\epair$, $\mupair$, $\taupair$ and 
inclusive hadronic final states and forward--backward asymmetries for
leptonic final states were similar to the ones performed at 
lower energies and the details, such as the event selection, and the
determination of the reduced energy $(\sqrt{\spr})$ can be found 
in~\cite{ref:delphi_130_172}, changes to each of the analyses are discussed 
below.

\begin{figure}[p]
\begin{center}
\begin{tabular}{cc}
 \multicolumn{2}{c}
 {\mbox{\epsfig{file=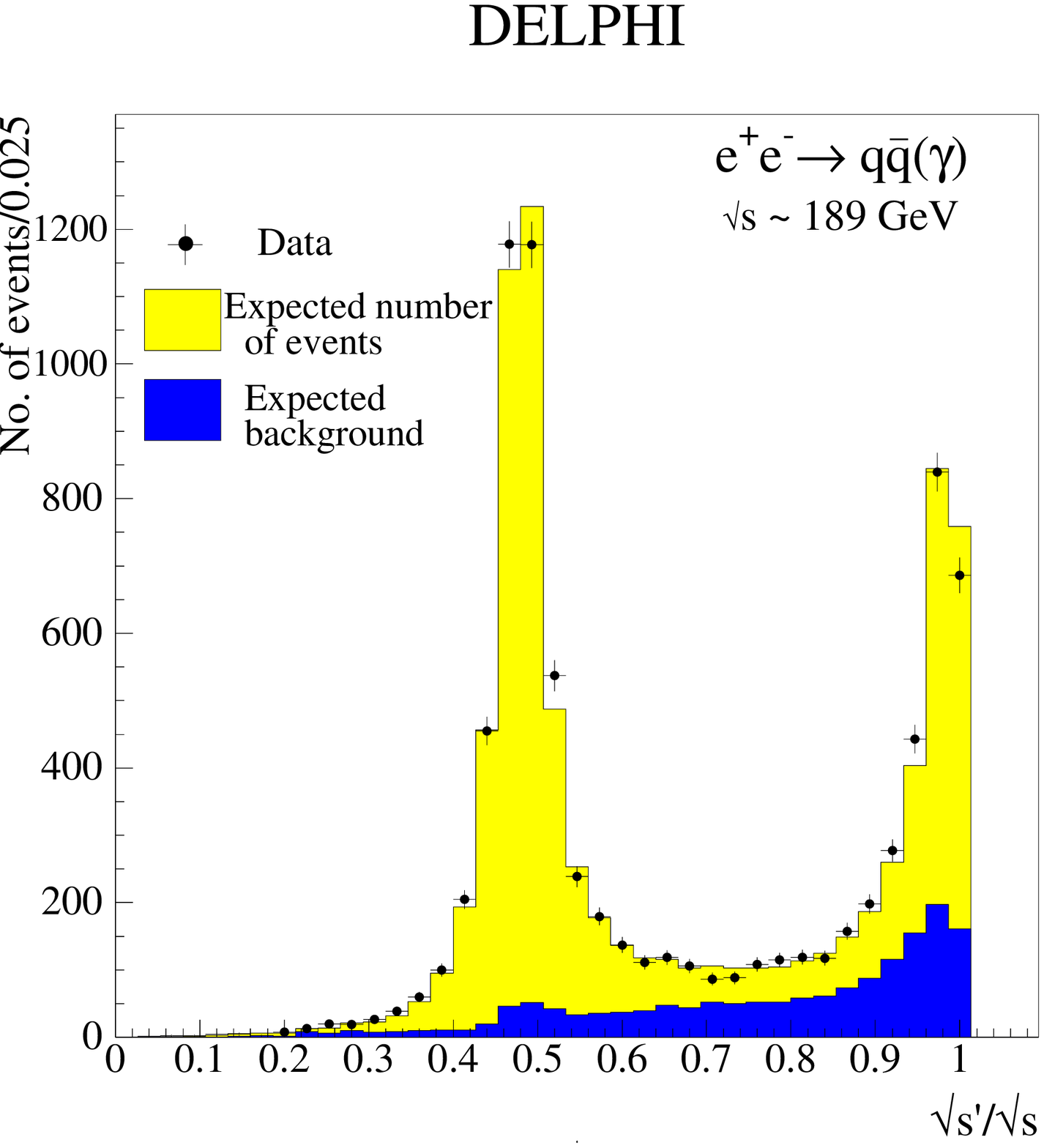,width=0.48\textwidth}}} \\
 \mbox{\epsfig{file=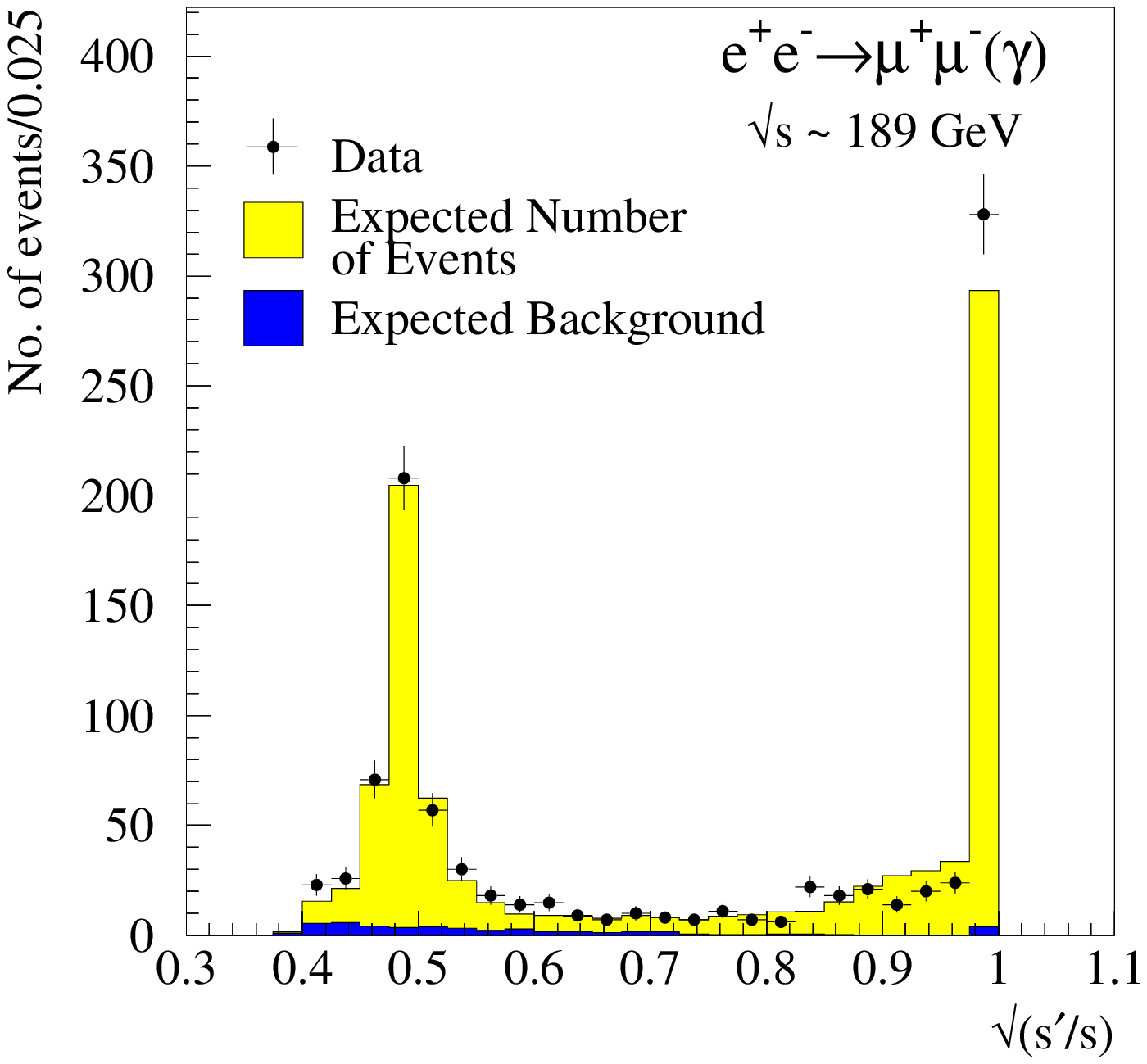,width=0.48\textwidth}} & 
 \mbox{\epsfig{file=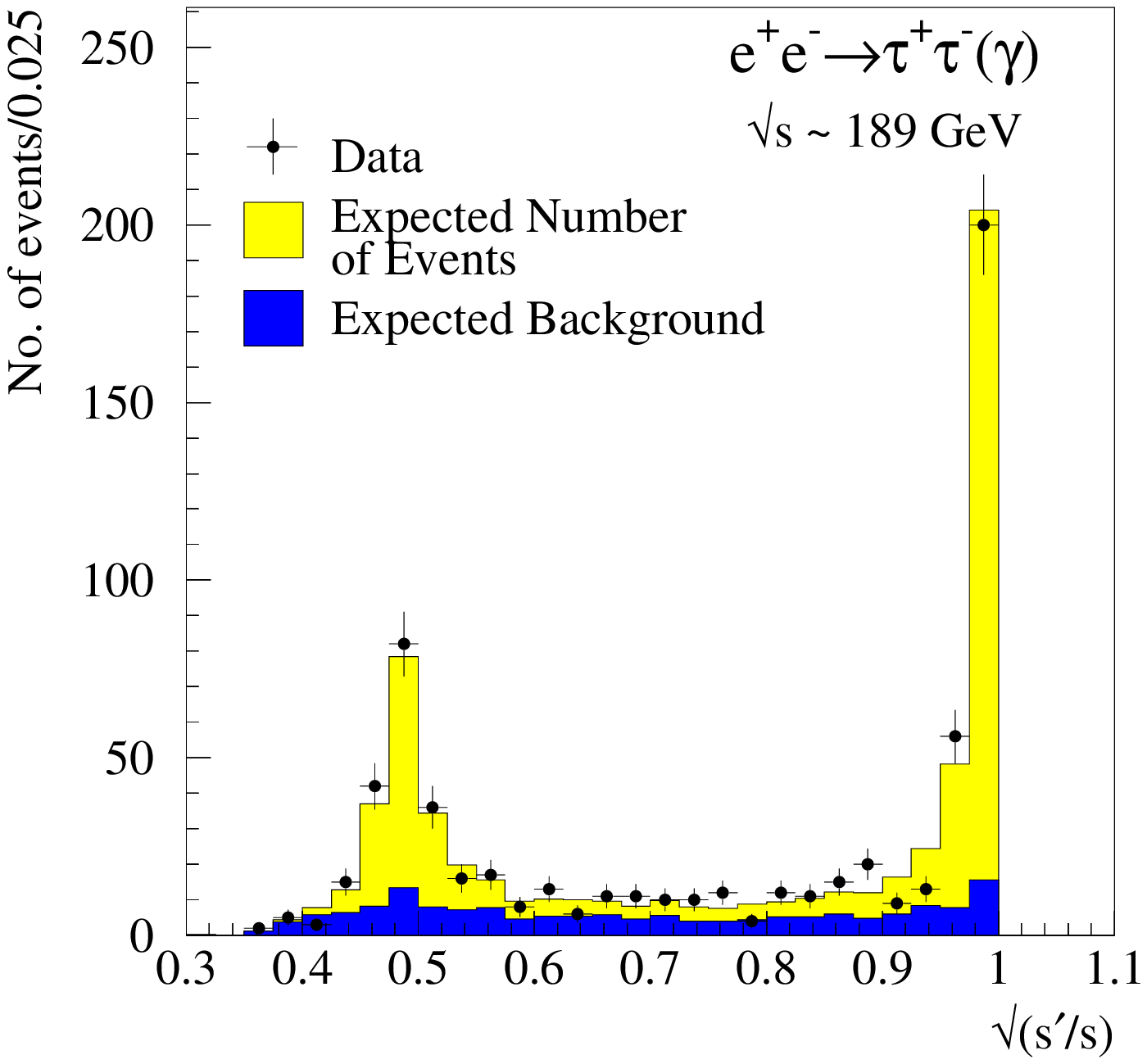,width=0.48\textwidth}} \\
\end{tabular}
\caption[]{\capsty 
{Distributions of the reconstructed reduced energy for the 
$\eemumug$, $\eettg$ and $\eeqqg$ processes at $\sqrt{s} \sim 189$ GeV. 
The points stand for the data and the histograms represent the signal
and background. The expected signals are simulated with the 
$\KORALZ$ generator~\cite{ref:koralz} for the $\eemumug$ and $\eettg$,
and with PYTHIA generator~\cite{ref:pythia} for the $\eeqqg$ channel.
The generator predictions were scaled to the 
$\ZFITTER$ \cite{ref:zfitter} predictions for the total cross-sections 
and are normalised to the luminosities of the data sets analysed.}}
\label{fig-SPRIME}
\end{center}
\end{figure}

The distributions of $\sqsps$ 
obtained for the real and the simulated data are shown in 
Figure~\ref{fig-SPRIME} for the muon, tau and inclusive hadronic channels for
$\sqs \sim 189$ GeV. 

The numbers of events selected in the {\emph{inclusive}}
samples for each final state are given in Table~\ref{tab-nev}. The 
efficiencies, the backgrounds from other channels and the backgrounds 
due to feed--up from the {\emph{inclusive}} samples 
for the {\emph{non--radiative}} event samples for each final states,
are given in Table~\ref{tab-ALLEFF}. 
Results are given in section~\ref{sec:results}

\begin{table}[t]
\begin{center}
\renewcommand{\arraystretch}{1.5}
\begin{tabular}{|c|c|c|}
\hline
            & \multicolumn{2}{|c|}{Energy $(\GeV)$} \\
\cline{2-3}
 Channel    & $\sim 183$     & $\sim 189$     \\
\hline\hline
 $\eeqqg$   & 5806           & 15726          \\
 $\eeeeg$   & 1109           & 2804           \\
 $\eemumug$ & 354            & 974            \\
 $\eettg$   & 253            & 632            \\
\hline
\end{tabular}
\caption[]{\capsty 
{The numbers of events used in the analyses of 
the different final states. For each channel, the values refer 
to the samples with $\sqrt{\spr/s} > 0.10$ for hadrons,
$\sqrt{\spr} > 75$ for muon and tau pairs and 
$\acol < 20 \mydeg$ for electron--positron pairs.}}
\label{tab-nev}
\end{center}
\end{table}

\begin{table}[p]
\begin{center}
\renewcommand{\arraystretch}{1.5}
\begin{tabular}{|c|c|c|c|c|}
\hline
            &   Energy   & Efficiency &  Background & Feed-up  \\
 Channel    &  ($\GeV$)  &   ($\%$)   &    ($\%$)   &  ($\%$)  \\
\hline    
\hline
 $\eeqq$    & $\sim 183$ &  92.6      &   30.6      & 10.5     \\
\cline{2-5}
            & $\sim 189$ &  92.6      &   33.2      & 10.5     \\ 
\hline
 $\eeee$    & $\sim 183$ &  96.8      &    0.2      &  0.8     \\ 
\cline{2-5}
            & $\sim 189$ &  97.9      &    0.2      &  0.8     \\ 
\hline
 $\eemm$    & $\sim 183$ &  89.0      &    1.0      &  1.8     \\ 
\cline{2-5}
            & $\sim 189$ &  92.3      &    0.5      &  2.2     \\ 
\hline
 $\eett$    & $\sim 183$ &  52.2      &   14.2      &  5.6     \\ 
\cline{2-5}
            & $\sim 189$ &  53.2      &   15.5      &  6.2     \\ 
\hline
\end{tabular}
\caption[]{\capsty
{The efficiency, backgrounds and feed--up from the inclusive samples
in the non--radiative samples of events selected in each of the channels.}}
\label{tab-ALLEFF}
\end{center}
\end{table}

\begin{table}[p]
\begin{center}
\renewcommand{\arraystretch}{1.5}
\begin{tabular}{|c|c|c|c|c|c|c|c|c|}
\hline
                                        &
 $\sqs$                                 &
 $\Delta\sigma^{h}/\sigma^{h}$          &
 $\Delta\sigma^{e}/\sigma^{e}$          &   
 $\Delta\sigma^{\mu}/\sigma^{\mu}$      &   
 $\Delta\sigma^{\tau}/\sigma^{\tau}$    &
 $\Delta\Afbe$                          &
 $\Delta\Afbm$                          &
 $\Delta\Afbt$                          \\
 & $\GeV$ & \% & \% & \% & \% &$ 10^{-3}$ & $10^{-3}$ & $10^{-3}$ \\

\hline\hline

{\it{Non-radiative}} & $\sim 183$ & 1.6 & 1.0 & 2.5 & 2.9 & $^{+10}_{-3}$ &  4 & 16 \\
                     & $\sim 189$ & 1.8 & 1.0 & 1.7 & 3.0 & $^{+10}_{-3}$ &  3 & 15 \\

\hline\hline

{\it{Inclusive}}     & $\sim 183$ & 1.1 & --  & 2.5 & 3.8 & --            &  2 & 16 \\
                     & $\sim 189$ & 1.1 & --  & 1.4 & 4.1 & --            &  2 & 15 \\
\hline

\end{tabular}
\caption[]{\capsty
{Systematic uncertainties of the total and non--radiative 
cross--section and forward--backward asymmetry measurements 
for the different final states. ``Non--radiative'' refers to
$\sqsps > 0.85$ for muon, tau and hadronic final states, 
and $\acol <20\mydeg$ for electron--positron pairs. ``Inclusive'' refers 
to $\sqsps > 0.10$ for the hadronic final states and 
to $\sqsp > 75~\GeV$ for the muon and tau final states.}}
\label{tab-ALLSYST}
\end{center}
\end{table}

\subsubsection{Inclusive hadronic final states}

A new cut was added. Events were only selected if their total
transverse energy\footnote{The total transverse energy is defined as
$E_{T} = \sum E_{i} \left| \sin \theta_{i} \right|$ where $E_{i}$ is 
the energy and $\theta_{i}$ is the polar angle of the $i^{\mathrm{th}}$ 
particle in the event. DELPHI uses a right handed coordinate system in 
which the $z$ axis is in the direction of the incoming electrons.}
was measured to be greater than $20\%$ of the collision energy.
This cut improves the rejection of two--photon collisions.
The sum of the energies of the charged particles in an event
was now required to be greater than $10\%$ of the 
collision energy, relaxing the cut from the previous analysis.

The selection efficiencies and backgrounds were determined 
from simulated events. The four--fermion 
background was determined from events generated by PYTHIA \cite{ref:pythia} 
and EXCALIBUR \cite{ref:excalibur}. The size of the background predicted by
the two generators was found to be in good agreement; residual differences
were taken into account in the systematic uncertainty on the measurement.
The main background contributions to the cross--section measurement at 183
(189) GeV came from  W-pair production with a contribution of 
$13.7 \pm 0.3 \: \pbarn$ ($14.7 \pm 0.3 \: \pbarn$) to the total 
cross--section and $8.3 \pm 0.2 \: \pbarn$ ($8.9 \pm 0.2 \: \pbarn$) to the
non--radiative cross--section. The combined production of Z--pair and
$\mathrm{Z}e^{+}e^{-}$ events was expected to contribute 
$2.8\pm 0.5 \: \pbarn$ ($3.4 \pm 0.5 \: \pbarn$) to the total cross--section
and $0.8 \pm 0.3 \: \pbarn$ ($1.1 \pm 0.2 \: \pbarn$) to the non--radiative
event sample. Using samples of events generated with the 
TWOGAM~\cite{ref:twogam} and BDKRC~\cite{ref:bdkrc} generators, two--photon 
collisions were found to contribute significantly only to the total 
cross--section measurement for which 
there remains $1.8 \pm 0.2 \: \pbarn$ after event selection cuts at 
both 183 and 189 \GeV.

\subsubsection{${\boldmath{e}}^{+}{\boldmath{e}}^{-}$ final states}

In \cite{ref:delphi_130_172} results were presented for two different cuts
on the acollinearity angle between the electron and positron
, $\acol < 90\mydeg$ and $\acol < 20\mydeg$. In this paper, 
results are given for $\acol < 20\mydeg$ 
only, which is the sample with the highest sensitivity to the models of
physics beyond the Standard Model considered in this paper.

\subsubsection{${\boldmath{\mu}}^{+}{\boldmath{\mu}}^{-}$ final states}

Two improvements were made for the analysis of the data 
collected at $\sqs \sim 189$ \GeV. The significant increase in
the luminosity at this energy made it possible to measure the efficiency of
the track reconstruction and muon identification efficiency from the data, 
exploiting the nearly back--to--back topology of the events with
$\sqsps > 0.85$ rather than relying on the efficiency determined from 
simulated events.

To do this, a sample of events with a high momentum muon was selected.
The efficiency was then determined from the number of these events which
did, or did not, contain a second reconstructed track or identified muon.
The uncertainty on the combined track reconstruction
and muon identification efficiency determined in this manner was $\pm 1.0\%$.
The difference between the efficiency determined directly from simulation 
and that derived from the data at $\sqs \sim 189$ \GeV\ was $\sim 2\% $.
The efficiency determined from the simulation at $\sqs \sim 183$ \GeV\
was corrected down by the $\sim 2\%$ difference measured at 189, and a 
systematic uncertainty of $\pm 2\%$ was applied to the results at 
$\sqs \sim 183$ \GeV.

The polar angular coverage was extended from $20 \mydeg < \theta < 160 \mydeg$ 
to $14 \mydeg < \theta < 164\mydeg$, taking advantage of the increased 
luminosity to perform checks of the tracking and muon identification 
efficiency at extreme polar angles.

The background coming from four--fermion final states,
via $\WWpair$, $\ZZpair$ and $\Zgst$ production, was estimated from events 
generated by EXCALIBUR. The background from two--photon collisions was 
estimated from events generated using BDKRC.

\subsubsection{${\boldmath{\tau}}^{+}{\boldmath{\tau}}^{-}$ final states}

There were several small improvements to the analysis of the $\eettg$ 
process. The acollinearity cut was placed at $0.3\mydeg$ instead of 
$0.5\mydeg$. Events with less than or equal to three charged 
particles in each hemisphere were included in the event selection 
( in addition to the events with only one charge particle in one 
hemisphere and less less than six charged particles in the other hemisphere )
provided that the reconstructed invariant mass in each
hemisphere was less than 2~\GeV$/c^{2}$, consistent with being the decay 
products of $\tau$ lepton. Both these changes improve the efficiency for the
signal while not significantly increasing the levels of background. 

The backgrounds were, as far as possible, 
estimated from the data by studying samples of events failing the 
specific cuts designed for rejection of a given background final state.
The total systematic uncertainties on the cross--section and forward--backward
asymmetry measurements for the different collision energies and channels are
shown in Table~\ref{tab-ALLSYST}.

In the determination of the forward--backward charge asymmetry of the
$\tau$ leptons, the scattering angle was taken as the polar angle of the 
highest momentum charged particle in the hemisphere determined to have 
originated from the negatively charged $\tau$ lepton. 
The asymmetry was corrected for acceptance, background and for 
contamination due to radiative events from lower $\sqsp$ values. 
The determined asymmetries and the associated uncertainties
are given in Tables~\ref{tab-xsec_afb} and \ref{tab-ALLSYST}
for the different centre-of-mass energies.

\subsection{Differential cross--sections}

In addition to the measurements of the cross--sections and asymmetries,
measurements of the differential cross--sections, $\dsdcth$,
are given for the $\mumu$ and $\tautau$ final states for the 
{\it{non--radiative}} samples.

\begin{figure}[p]
\begin{center}
\mbox{\epsfig{file=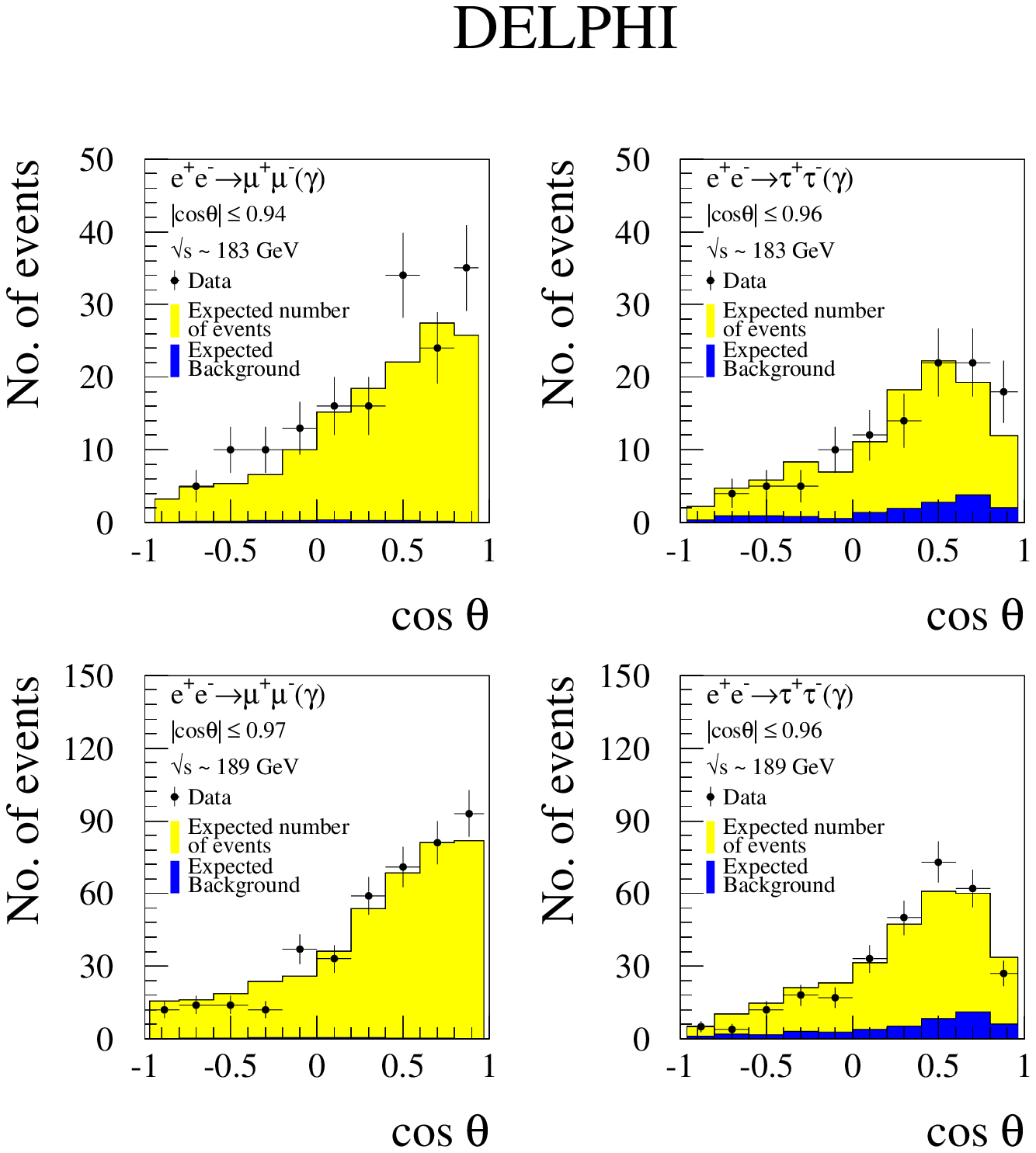,width=\textwidth}} 
\caption[]{\capsty{The numbers of events observed as a function of 
$\cos \theta$ for $\mumu$ and $\tautau$ final states at centre--of--mass 
energies of $\sim 183$ and 189 GeV. 
The points stand for the data and the histograms represent the signal
and background. The expected signals are simulated with the 
$\KORALZ$ \cite{ref:koralz} generator scaled to the 
$\ZFITTER$ \cite{ref:zfitter} predictions and normalised to the luminosities
of the data sets analysed.}}
\label{fig-diffxs}
\end{center}
\end{figure}

Figure \ref{fig-diffxs} shows the numbers of events
observed in bins of $\cos\theta$ compared to simulations for each final state
and each collision energy.
For the $\mumu$ final states the scattering angle $\theta$ is the 
angle of the negative fermion with respect to the incoming electron in 
the laboratory frame, for the $\tautau$ final states the angle was defined 
in the same way as mentioned above for the measurement of the forward--backward
asymmetry. Results are given in section~\ref{sec:results}

\subsection{Results of analyses}
\label{sec:results}

The results of the cross--section and asymmetry measurements are presented in 
Table~\ref{tab-xsec_afb} together with theoretical predictions. 
The errors indicated are statistical only. 
Systematic errors due to the event selection and to
the residual background subtraction are shown in 
Table~\ref{tab-ALLSYST}. 
For the cross--section measurements, they must be added in quadrature to 
the uncertainty coming from the luminosity determination.
The theoretical predictions in Table~\ref{tab-xsec_afb}
are from the $\TOPAZZERO$ program~\cite{ref:topazzero} for
electron--positron final states and $\ZFITTER$ program~\cite{ref:zfitter} 
for the other final states. The uncertainties on the theoretical 
predictions are estimated to be below $1\%$.

Some components of the systematic uncertainties are correlated between 
measurements  in different channels and different energies. 
This is the case for the theoretical
uncertainty on the luminosity determination which is correlated between all
cross-section measurements at all energies. For the \mumu\ final states,
the uncertainty on the event selection efficiency is correlated between
energy points. The uncertainty on the extrapolation to $4\pi$ acceptance
coming from the interference between initial and final state radiation is
correlated between \mumu\ and \tautau\ final states and between energies.
Given the estimated size of the correlations compared to the
precision of the measurements these correlations are ignored.

\begin{figure}[p]
\begin{center}
\begin{tabular}{c}
\mbox{\epsfig{file=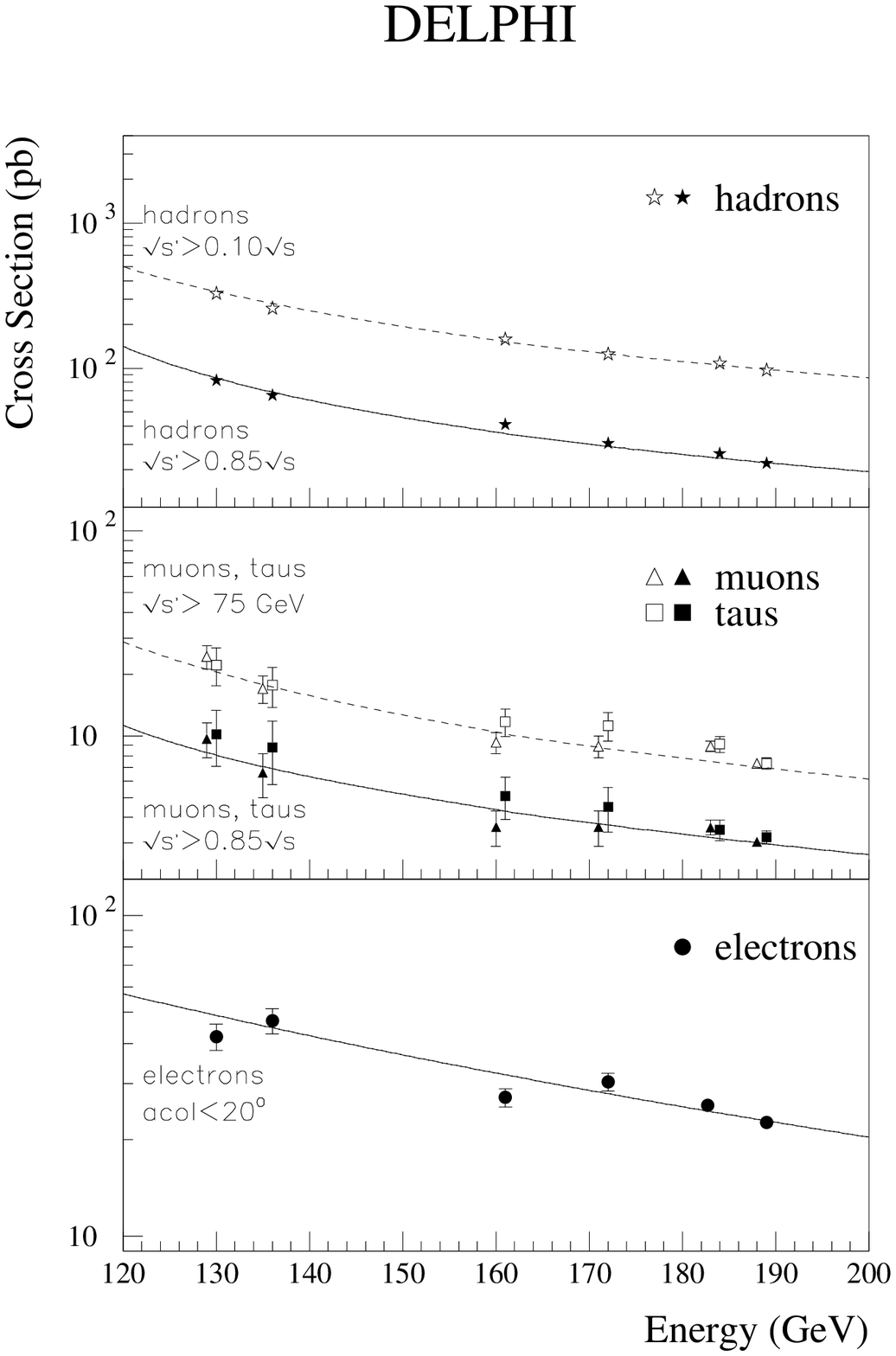,width=0.80\textwidth}} \\
\end{tabular}
\caption[]{\capsty 
{Cross--sections for the $\eeqqg$, $\mumug$ and $\tautaug$ and $\eeeeg$
processes measured at energies from 130 up to 189~$\GeV$. The curves show
the SM prediction of the $\TOPAZZERO$ program~\cite{ref:topazzero} for
electron--positron final states and $\ZFITTER$
program~\cite{ref:zfitter} for the other final states. 
Solid points and solid lines represent the {\emph{non--radiative}} selections,
open points and dashed lines represent the {\emph{inclusive}} selections.}}
\label{fig-CS}
\end{center}
\end{figure}

\begin{figure}[p]
\begin{center}
\begin{tabular}{c}
\mbox{\epsfig{file=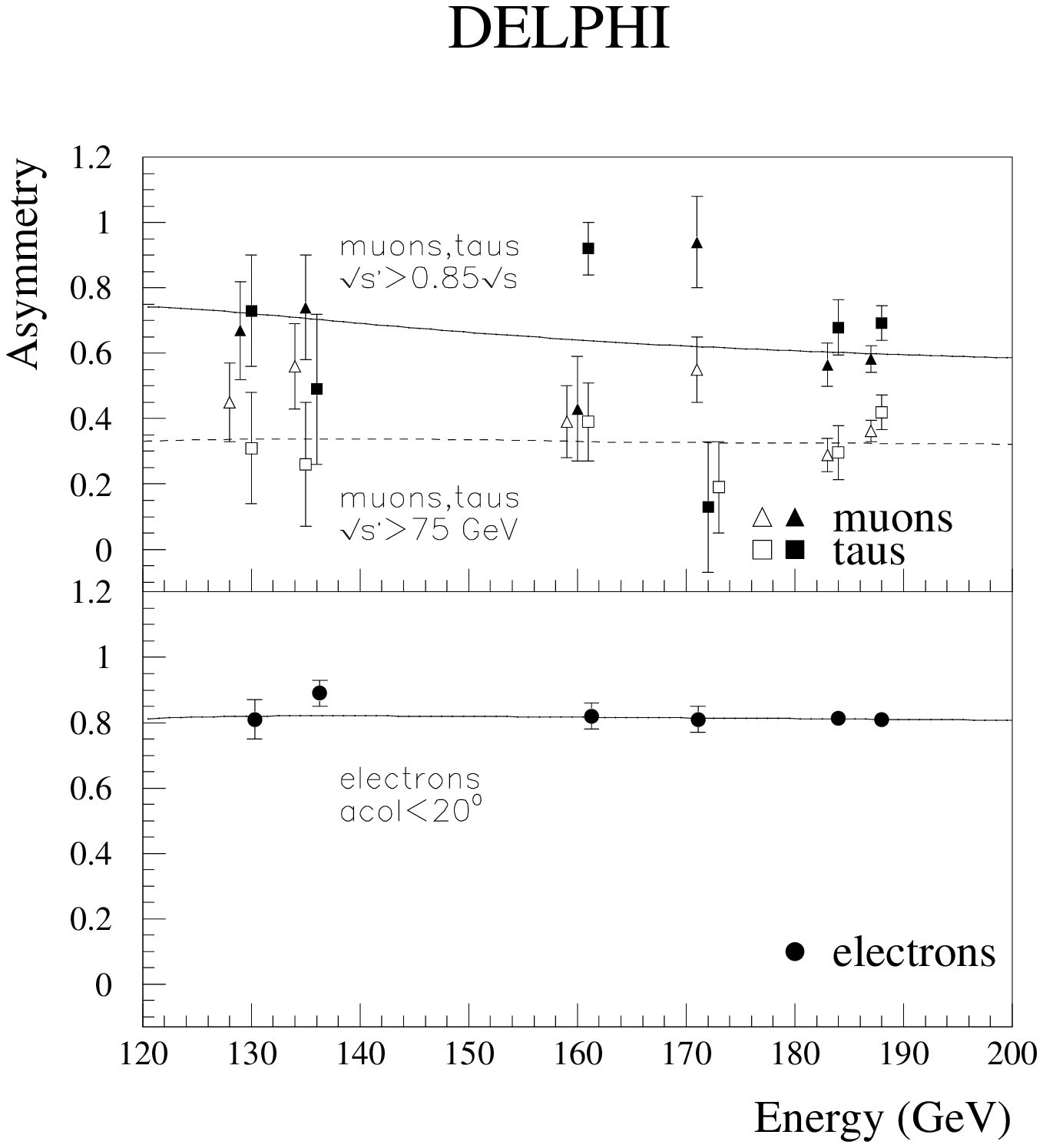,width=\textwidth}} \\
\end{tabular}
\caption[]{\capsty 
{The forward--backward charge asymmetries in the reactions 
$\eemumug$, $\tautaug$ and $\eeg$ measured at energies ranging from 
130 to 189 GeV. The curves show
the SM prediction of the $\TOPAZZERO$ program~\cite{ref:topazzero} for
electron--positron final states and $\ZFITTER$ 
program~\cite{ref:zfitter} for the other final states.
Solid points and solid lines represent the {\emph{non--radiative}} selections,
open points and dashed lines represent the {\emph{inclusive}} selections.}}
\label{fig-AS}
\end{center}
\end{figure}

Figure~\ref{fig-CS} shows the measured hadron, electron--positron pair,
muon--pair and tau--pair cross--sections
for all collision energies ranging from 130 up to 189~$\GeV$ from DELPHI. 
The forward--backward asymmetries for electron--positron pairs, 
muon--pairs and tau--pairs are shown
in figure \ref{fig-AS}.

The results of the analyses of the differential cross--sections for 
$\mumu$ and $\tautau$ final states are tabulated, including statistical and 
systematic errors, in Table \ref{tab-diffxs}. The theoretical predictions 
are from the ZFITTER program, and have an uncertainty estimated to be 
below $1\%$. 

For the $\eemm$ channel the systematic errors quoted in 
Table~\ref{tab-diffxs} include correlated systematic 
uncertainties of $2 \%$ for the data at $\sqs \sim 183 \: \GeV$ and 
$1 \%$ for the data at $\sqs \sim 189 \: \GeV$ in the measured
cross--sections for all bins of $\cos\theta$ arising from the
determination of the track reconstruction and muon identification 
efficiencies which were applied as an overall correction to the
efficiencies determined bin by bin in $\cos\theta$ from simulated
events.

Overall, no substantial departure of the measurements of fermion--pair
production from the Standard Model predictions was found.

\begin{table}[p]
\begin{center}
\renewcommand{\arraystretch}{1.2}
\begin{tabular}{|c|r|c|c|}
\hline
\multicolumn{2}{|c|}{Energy $(\GeV)$}  
                                       & $\sim 183$        & $\sim 189$        \\
\hline\hline
$\sigma_{had}(pb)$ & $\sqsps>0.85$     & $25.8 \pm 0.8$    & $ 22.1 \pm 0.5$   \\
                   & Theory            &     23.8          &      21.7         \\
\cline{2-4}
                   & $\sqsps>0.10$     & $107.8 \pm 1.7$   & $ 97.1 \pm 1.0$   \\
                   & Theory            &     106.0         &      97.4         \\
\hline\hline
$\smu (pb)$        & $\sqsps>0.85$     & $3.58 \pm 0.28$   & $3.04 \pm 0.15$   \\
                   & Theory            &     3.31          &      3.08         \\
\cline{2-4}
                   & $\sqsp>75~\GeV$   & $8.92 \pm 0.47$   & $7.34 \pm 0.24$   \\
                   & Theory            &     7.69          &      7.15         \\
\hline
$\stau (pb)$       & $\sqsps>0.85$     & $3.48 \pm 0.39$   & $3.21 \pm 0.22$   \\
                   & Theory            &     3.39          &      3.16         \\
\cline{2-4}
                   & $\sqsp>75~\GeV$   & $9.12 \pm 0.73$   & $7.35 \pm 0.39$   \\
                   & Theory            &     7.73          &      7.18         \\
\hline\hline
$\Afbm$            & $\sqsps>0.85$     & $0.565 \pm 0.067$ & $0.582 \pm 0.041$ \\
                   & Theory            &     0.594         &      0.588        \\
\cline{2-4}
                   & $\sqsp>75~\GeV$   & $0.289 \pm 0.051$ & $0.362 \pm 0.032$ \\
                   & Theory            &     0.317         &      0.308        \\
\hline
$\Afbt$            & $\sqsps>0.85$     & $0.679 \pm 0.082$ & $0.693 \pm 0.051$ \\
                   & Theory            &     0.594         &      0.587        \\
\cline{2-4}
                   & $\sqsp>75~\GeV$   & $0.296 \pm 0.081$ & $0.420 \pm 0.050$ \\
                   & Theory            &     0.316         &      0.315        \\
\hline
\hline
$\see (pb)$        & $\acol<20\mydeg$  & $25.6 \pm 0.8$    & $22.6 \pm 0.4$    \\
                   & Theory            &     24.7          &      23.1         \\
\hline\hline
$\Afbe$            & $\acol<20\mydeg$  & $0.814 \pm 0.017$ & $0.810 \pm 0.010$ \\
                   & Theory            &     0.820         &      0.821        \\
\hline
\end{tabular}
\caption[]{\capsty 
{Results of the cross--section and asymmetry measurements for 
the different final states. The errors indicated are 
statistical only. Systematic errors related to the event selection 
and residual backgrounds are provided in Table~\ref{tab-ALLSYST}.
Those coming from the luminosity determination are given in   
the text. 
The hadronic, muon and tau results are corrected for all cuts, apart 
from the $\sqrt{s^{\prime}}$ cut.}}
\label{tab-xsec_afb}
\end{center}
\end{table}

\begin{table}[p]
\begin{center}
\renewcommand{\arraystretch}{1.5}
\begin{tabular}{c@{\hspace{.5cm}}c}
\begin{tabular}{|@{$\: [$}r@{,}r@{$] \:$}|c|rr@{$\pm$}c@{$\pm$}c|}
\hline
\multicolumn{7}{|c|}{\boldmath{$\eemm \: (\sqs \sim 183 \: \GeV)$}} \\
\hline\hline
\multicolumn{2}{|c|}{} & \multicolumn{5}{|c|}{$ \dsdcth \: (\upb)$} \\
\cline{3-7}
\multicolumn{2}{|c|}{$\cos\theta$} & Theory & 
                                  \multicolumn{4}{|c|}{Measurement} \\
\hline\hline
-0.94 & -0.80 & 0.478 &   & 0.000 & 0.178 & 0.013 \\
-0.80 & -0.60 & 0.486 &   & 0.514 & 0.230 & 0.013 \\
-0.60 & -0.40 & 0.576 &   & 0.989 & 0.313 & 0.024 \\
-0.40 & -0.20 & 0.761 &   & 0.972 & 0.307 & 0.023 \\
-0.20 &  0.00 & 1.045 &   & 1.298 & 0.360 & 0.032 \\
 0.00 &  0.20 & 1.428 &   & 1.591 & 0.398 & 0.039 \\
 0.20 &  0.40 & 1.913 &   & 1.605 & 0.401 & 0.039 \\
 0.40 &  0.60 & 2.503 &   & 3.377 & 0.579 & 0.081 \\
 0.60 &  0.80 & 3.206 &   & 2.466 & 0.503 & 0.061 \\
 0.80 &  0.94 & 4.078 &   & 4.978 & 0.841 & 0.119 \\
\hline
\end{tabular} 
&
\begin{tabular}{|@{$\: [$}r@{,}r@{$] \:$}|c|r@{}r@{$\pm$}c@{$\pm$}c|}
\hline
\multicolumn{7}{|c|}{\boldmath{$\eett \: (\sqs \sim 183 \: \GeV)$}} \\
\hline\hline
\multicolumn{2}{|c|}{} & \multicolumn{5}{|c|}{$ \dsdcth \: (\upb)$} \\
\cline{3-7}
\multicolumn{2}{|c|}{$\cos\theta$} & Theory &
                                  \multicolumn{4}{|c|}{Measurement} \\
\hline\hline
-0.96 & -0.80 & 0.49  & - & 0.13 & 0.24 & 0.04 \\
-0.80 & -0.60 & 0.50  &   & 0.48 & 0.31 & 0.04 \\
-0.60 & -0.40 & 0.60  &   & 0.52 & 0.28 & 0.04 \\
-0.40 & -0.20 & 0.79  &   & 0.56 & 0.30 & 0.04 \\
-0.20 &  0.00 & 1.08  &   & 1.62 & 0.54 & 0.12 \\
 0.00 &  0.20 & 1.48  &   & 1.56 & 0.51 & 0.12 \\
 0.20 &  0.40 & 1.97  &   & 1.65 & 0.51 & 0.12 \\
 0.40 &  0.60 & 2.58  &   & 2.49 & 0.61 & 0.19 \\
 0.60 &  0.80 & 3.31  &   & 3.91 & 1.00 & 0.29 \\
 0.80 &  0.96 & 4.08  &   & 6.77 & 1.80 & 0.50 \\
\hline
\end{tabular}
\\ \\
\begin{tabular}{|@{$\: [$}r@{,}r@{$] \:$}|c|rr@{$\pm$}c@{$\pm$}c|}
\hline
\multicolumn{7}{|c|}{\boldmath{$\eemm \: (\sqs \sim 189 \: \GeV)$}} \\
\hline\hline
\multicolumn{2}{|c|}{} & \multicolumn{5}{|c|}{$ \dsdcth \: (\upb)$} \\
\cline{3-7}
\multicolumn{2}{|c|}{$\cos\theta$} & Theory &
                                  \multicolumn{4}{|c|}{Measurement} \\
\hline\hline
-0.97 & -0.80 & 0.465 &   & 0.495 & 0.143 & 0.008 \\
-0.80 & -0.60 & 0.467 &   & 0.478 & 0.128 & 0.008 \\
-0.60 & -0.40 & 0.546 &   & 0.448 & 0.120 & 0.007 \\
-0.40 & -0.20 & 0.713 &   & 0.391 & 0.113 & 0.006 \\
-0.20 &  0.00 & 0.971 &   & 1.287 & 0.212 & 0.021 \\
 0.00 &  0.20 & 1.322 &   & 1.129 & 0.197 & 0.018 \\
 0.20 &  0.40 & 1.769 &   & 1.908 & 0.248 & 0.029 \\
 0.40 &  0.60 & 2.315 &   & 2.445 & 0.290 & 0.039 \\
 0.60 &  0.80 & 2.968 &   & 2.927 & 0.325 & 0.048 \\
 0.80 &  0.97 & 3.780 &   & 3.986 & 0.413 & 0.065 \\
\hline
\end{tabular}
&
\begin{tabular}{|@{$\: [$}r@{,}r@{$] \:$}|c|rr@{$\pm$}c@{$\pm$}c|}
\hline
\multicolumn{7}{|c|}{\boldmath{$\eett \: (\sqs \sim 189 \: \GeV)$}} \\
\hline\hline
\multicolumn{2}{|c|}{} & \multicolumn{5}{|c|}{$ \dsdcth \: (\upb)$} \\
\cline{3-7}
\multicolumn{2}{|c|}{$\cos\theta$} & Theory &
                                  \multicolumn{4}{|c|}{Measurement} \\
\hline\hline
-0.96 & -0.80 & 0.48  &   & 0.58 & 0.34 & 0.05 \\
-0.80 & -0.60 & 0.48  &   & 0.12 & 0.13 & 0.03 \\
-0.60 & -0.40 & 0.56  &   & 0.48 & 0.16 & 0.04 \\
-0.40 & -0.20 & 0.73  &   & 0.67 & 0.19 & 0.05 \\
-0.20 &  0.00 & 1.00  &   & 0.75 & 0.22 & 0.06 \\
 0.00 &  0.20 & 1.37  &   & 1.57 & 0.31 & 0.13 \\
 0.20 &  0.40 & 1.83  &   & 2.05 & 0.32 & 0.16 \\
 0.40 &  0.60 & 2.39  &   & 2.96 & 0.39 & 0.23 \\
 0.60 &  0.80 & 3.06  &   & 3.26 & 0.51 & 0.26 \\
 0.80 &  0.96 & 3.78  &   & 2.87 & 0.71 & 0.24 \\
\hline
\end{tabular}
\\
\end{tabular}
\end{center}
\caption[]{\capsty{The differential cross--sections for non--radiative
$\mumu$ and $\tautau$ final states at centre--of--mass energies of 
$\sim 183$ and 189 GeV. The errors shown are respectively the statistical and
systematic components. The Standard Model expectations (SM) were 
computed with the ZFITTER program \cite{ref:zfitter}.}}
\label{tab-diffxs}
\end{table}

\section{Physics beyond the Standard Model}
\label{sec:interpretation}


The data presented in this paper were used to improve the constraints on 
physics beyond the Standard Model given in section 6 of
\cite{ref:delphi_130_172} for three sets of models: contact interactions 
between leptons, models including Z$^\prime$ bosons
and R-parity violating sneutrino exchange. The theoretical bases of each of 
these models are discussed in section 5 of \cite{ref:delphi_130_172}, the key
points are summarised below.
New limits for models which include gravity in extra dimensions are derived
from the measurements of the differential cross--sections given in this paper.
Unless otherwise stated the systematic errors on the measurements at 
LEP II energies have been added in quadrature with the statistical errors
treating them as uncorrelated between measurements.

\subsection{Contact interaction models}

Contact interactions between fermions can be parameterised as an effective 
Lagrangian with the form: 
\begin{equation}
 \mbox{$\cal{L}$}_{eff} = \frac{g^{2}}{(1+\delta)\Lambda^{2}} 
                          \sum_{i,j=L,R} \eta_{ij} 
                           \overline{e}_{i} \gamma_{\mu} e_{i}
                            \overline{f}_{j} \gamma^{\mu} f_{j}.
 \label{eqn:cntclag}
\end{equation}
where $\Lambda$ is the characteristic energy scale of the interactions.
Different choices of $\eta_{ij}$ lead to 12 commonly studied models,
referred to as LL, RR etc~\cite{ref:cntc-thry}. 

Fits were made using data at all energies from 130 to 189 GeV
for $\eeee, \eemm, \eett$ channels and $\eell$, a
combination of all leptonic final states assuming lepton universality.
The parameter fitted was $\epsilon=1/\Lambda^{2}$.
The values of $\epsilon$ extracted for each model were all compatible 
with the Standard Model expectation $\epsilon=0$, at the two standard 
deviation level. The errors on $\epsilon$ in the $\eell$ fit are
typically $30\%$ smaller than those reported in 
\cite{ref:delphi_130_172} as a result of the inclusion of the data 
collected at $\sqrt{s} \sim 183$ and 189 $\GeV$. 
The fitted values of $\epsilon$ were 
converted into lower limits on $\Lambda$ at $95\%$ confidence level. 
The results are given in Table \ref{tab:llL}.

\begin{table}[p]
\vskip 4cm
 \begin{center}
 \renewcommand{\arraystretch}{1.5}
 \begin{tabular}{cc}
  \begin{tabular}{|c|r|c|c|}
   \hline
   \multicolumn{4}{|c|}{\boldmath $\eeee$ \unboldmath} \\
   \hline
   \hline
   Model  & $\epsilon^{+\sigma_{+}}_{-\sigma_{-}} (\TeV^{-2})$ &
  $\Lambda^{+} (\TeV)$ & $\Lambda^{-} (\TeV)$ \\
   \hline
   \hline
   LL &  0.016$^{+ 0.022}_{- 0.020}$ &    4.4 &    5.4 \\
   \hline
   RR &  0.016$^{+ 0.023}_{- 0.020}$ &    4.3 &    5.3 \\
   \hline
   VV &  0.002$^{+ 0.005}_{- 0.004}$ &    9.8 &   11.7 \\
   \hline
   AA &  0.007$^{+ 0.010}_{- 0.014}$ &    6.6 &    7.1 \\
   \hline
   RL &  0.003$^{+ 0.018}_{- 0.013}$ &    5.5 &    6.3 \\
   \hline
   LR &  0.003$^{+ 0.018}_{- 0.013}$ &    5.5 &    6.3 \\
   \hline
  \end{tabular}
  &
  \begin{tabular}{|c|r|c|c|}
   \hline
   \multicolumn{4}{|c|}{\boldmath $\eemm$ \unboldmath} \\
   \hline
   \hline
   Model  & $\epsilon^{+\sigma_{+}}_{-\sigma_{-}} (\TeV^{-2})$ &
  $\Lambda^{+} (\TeV)$ & $\Lambda^{-} (\TeV)$ \\
   \hline
   \hline
   LL & -0.002$^{+ 0.013}_{- 0.014}$ &    6.6 &    6.3 \\
   \hline
   RR & -0.002$^{+ 0.014}_{- 0.016}$ &    6.3 &    5.9 \\
   \hline
   VV &  0.001$^{+ 0.004}_{- 0.006}$ &   10.9 &   10.1 \\
   \hline
   AA & -0.003$^{+ 0.009}_{- 0.005}$ &    9.1 &    9.2 \\
   \hline
   RL & -0.252$^{+ 0.261}_{- 0.016}$ &    2.1 &    1.9 \\
   \hline
   LR & -0.252$^{+ 0.261}_{- 0.016}$ &    2.1 &    1.9 \\
   \hline
  \end{tabular}
  \\ \\
  \begin{tabular}{|c|r|c|c|}
   \hline
   \multicolumn{4}{|c|}{\boldmath $\eett$ \unboldmath} \\
   \hline
   \hline
   Model  & $\epsilon^{+\sigma_{+}}_{-\sigma_{-}} (\TeV^{-2})$ &
  $\Lambda^{+} (\TeV)$ & $\Lambda^{-} (\TeV)$ \\
   \hline
   \hline
   LL &  0.004$^{+ 0.020}_{- 0.022}$ &    5.2 &    5.4 \\
   \hline
   RR &  0.004$^{+ 0.023}_{- 0.023}$ &    4.9 &    5.1 \\
   \hline
   VV & -0.011$^{+ 0.009}_{- 0.006}$ &    9.0 &    7.0 \\
   \hline
   AA &  0.019$^{+ 0.012}_{- 0.009}$ &    5.1 &    7.8 \\
   \hline
   RL & -0.163$^{+ 0.100}_{- 0.049}$ &    2.9 &    2.0 \\
   \hline
   LR & -0.163$^{+ 0.100}_{- 0.049}$ &    2.9 &    2.0 \\
   \hline
  \end{tabular}
  &
  \begin{tabular}{|c|r|c|c|}
   \hline
   \multicolumn{4}{|c|}{\boldmath $\eell$ \unboldmath} \\
   \hline
   \hline
   Model  & $\epsilon^{+\sigma_{+}}_{-\sigma_{-}} (\TeV^{-2})$ &
  $\Lambda^{+} (\TeV)$ & $\Lambda^{-} (\TeV)$ \\
   \hline
   \hline
   LL &  0.005$^{+ 0.009}_{- 0.011}$ &    7.3 &    7.8 \\
   \hline
   RR &  0.004$^{+ 0.011}_{- 0.010}$ &    6.8 &    7.6 \\
   \hline
   VV &  0.001$^{+ 0.002}_{- 0.004}$ &   14.5 &   12.7 \\
   \hline
   AA &  0.006$^{+ 0.005}_{- 0.005}$ &    8.3 &   10.9 \\
   \hline
   RL & -0.008$^{+ 0.010}_{- 0.011}$ &    7.6 &    6.2 \\
   \hline
   LR & -0.008$^{+ 0.010}_{- 0.011}$ &    7.6 &    6.2 \\
   \hline
  \end{tabular}
 \\
 \end{tabular}
 \end{center}
 \caption{\capsty{ 
               {Fitted values of $\epsilon$
               and $95\%$ confidence lower limits on the scale,
               $\Lambda$, of contact interactions in the models 
               discussed in the text, for $\eeee$, $\eemm$, $\eett$ 
               final states 
               and also for $\eell$ in which lepton universality 
               is assumed for the contact interactions. 
               The errors on $\epsilon$ are statistical only. 
               The models are defined in \cite{ref:cntc-thry}.}}}
\label{tab:llL}
\vskip 4cm
\end{table}

\subsection{Sneutrino exchange}
The second set of models consider possible $s$ or $t$ channel sneutrino 
($\snul$) exchange in R-parity violating supersymmetry \cite{ref:rpsusy-thry}. 
The parameters of interest are the dimensionless couplings, 
$\lambda_{ijk}$, between the
superfields of different generations, $i,j$ and $k$, together with 
the mass of the sneutrino exchanged, $\msneut$. The sneutrino width is not 
constrained within R-parity violating supersymmetry; a value of 1 GeV has been 
used~\cite{ref:rpsusy-thry}. 

For the $\eemm$ and $\eett$ channels, in the case that only one $\lambda$
value is non--zero there would only be $t$-channel sneutrino effects.
The $95\%$ confidence exclusion upper limits on $\lambda$ are given in 
Table \ref{tab:susy}, assuming sneutrino masses of either 
100 or 200~$\GeV/c^{2}$. The limits are calculated by finding the value
of $\lambda$ for $\chi^{2} = \chi^{2}_{min} + 3.84$.
The limits are between 0.02 and 0.14 lower than those published in
\cite{ref:delphi_130_172} depending on the channel and the mass assumed.

\begin{figure}[p]
\vspace*{-2.0cm}
 \begin{center}
  \epsfig{file=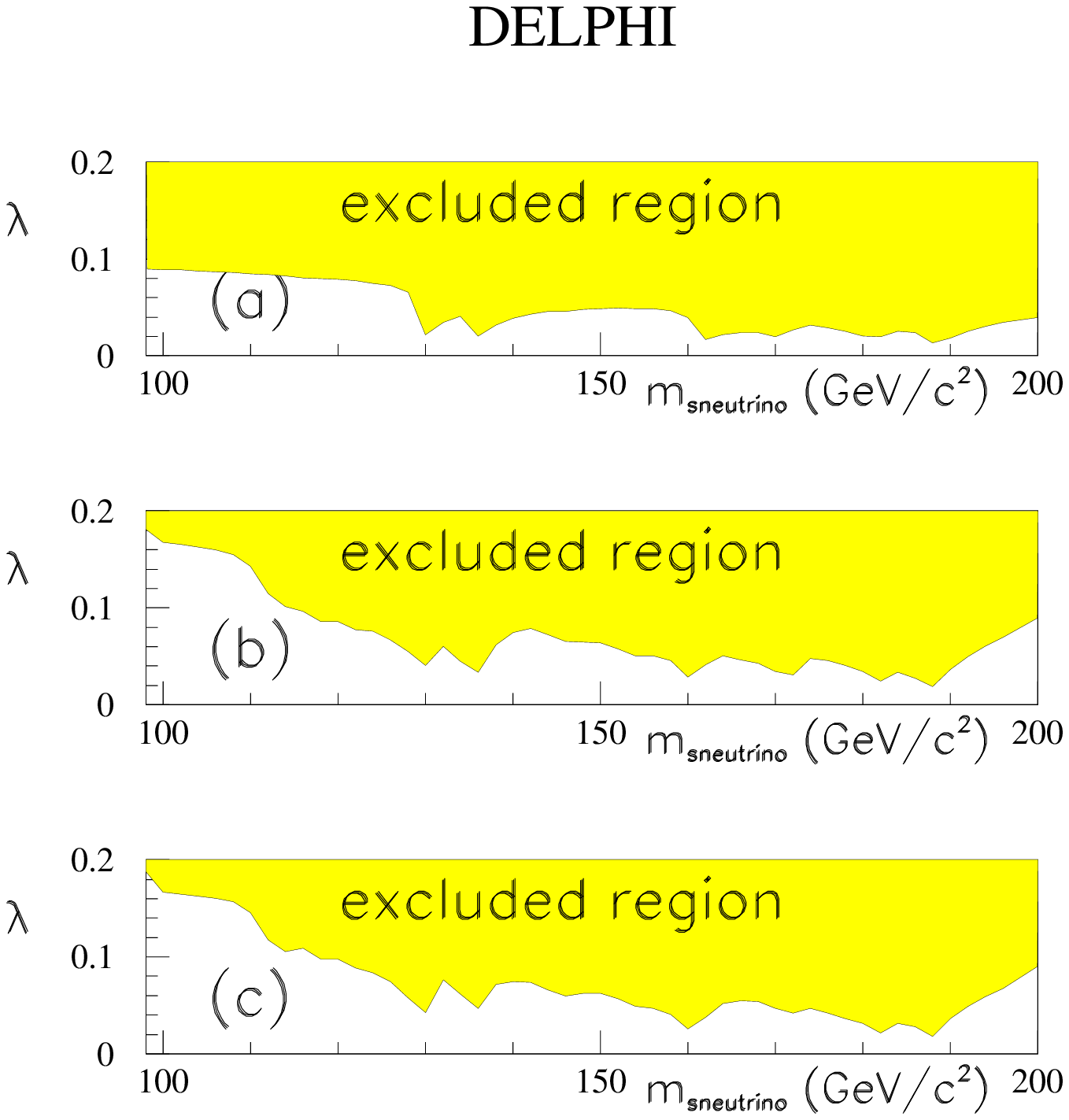,height=16.0cm,width=\textwidth}
 \end{center}
 \caption[]{\capsty
               {The $95\%$ exclusion limits
               for (a) $\lambda_{121}$ (or $\lambda_{131}$), as
               a function of $\msneut$, obtained from the $\eeee$ channel;
               (b) $\lambda_{131} = \lambda_{232} = \lambda$, as
               a function of $\msneut$, obtained from the $\eemumu$ channel;
               (c) $\lambda_{121} = \lambda_{233} = \lambda$, as
               a function of $\msneut$, obtained from the $\eetautau$ channel.
               The sneutrino width is taken to be 1 GeV. }}
 \label{fig:exclall}
\end{figure}

For the $\eeee$ channel the resulting $95\%$ limits on
$\lambda$, are given in Figure \ref{fig:exclall}(a), 
as a function of $\msneut$.
For the fits in the $\eemumu$ channel, assuming that 
$\lambda_{131} = \lambda_{232} = \lambda$, the resulting $95\%$ limits on 
$\lambda$ are given in Figure \ref{fig:exclall}(b).
The exclusion contour for $\lambda_{121} = \lambda_{233} = \lambda$, using
the $\eetautau$ channel, is shown in Figure \ref{fig:exclall}(c). 
In each case, the exclusion contours are calculated by finding the value
of $\lambda$ for $\chi^{2} = \chi^{2}_{min} + 3.84$ for each value
of $\msneut$ separately. A coupling
of $\lambda > 0.1$ can be excluded for $\msneut$ in the range 
130 - 190 $\GeV/c^{2}$ for all final states, extending the excluded region 
by approximately 20 $\GeV/c^{2}$ compared to \cite{ref:delphi_130_172}.

\begin{table}[p]
 \begin{center}
  \renewcommand{\arraystretch}{1.5}
  \begin{tabular}{|c|c|c|}
   \hline
                    & $\msneut=100 \: \GeV/c^2$ & $\msneut=200 \: \GeV/c^2$ \\
     coupling       &        ($95\%$ c.l.)       &        ($95\%$ c.l.)     \\
   \hline
   \hline
     $\lambda$ ($t$-chann. $\snul$ in $\eemumu$) &
           0.50            &           0.68            \\
   \hline
     $\lambda$ ($t$-chann. $\snul$ in $\eetautau$) &
           0.47            &           0.65            \\
   \hline
  \end{tabular}
 \end{center}
 \caption{\capsty
               {Upper limits on the couplings $\lambda$  in
               $t$ channel sneutrino exchange in $\eemumu$  and
               $\eetautau$ for sneutrino masses of 100 and 200 $\GeV/c^{2}$.
               The couplings involved are given in the text.}}
 \label{tab:susy}
\end{table}

\subsection{\boldmath{$\Zprime$}-bosons}

Existing data from LEP1 and LEP2 and the cross--sections and 
asymmetries given here were used to fit the data to models including
additional $\Zprime$ bosons.

\subsubsection{Model dependent fits}

\begin{figure}[p]
 \begin{center}
   \mbox{\epsfig{file=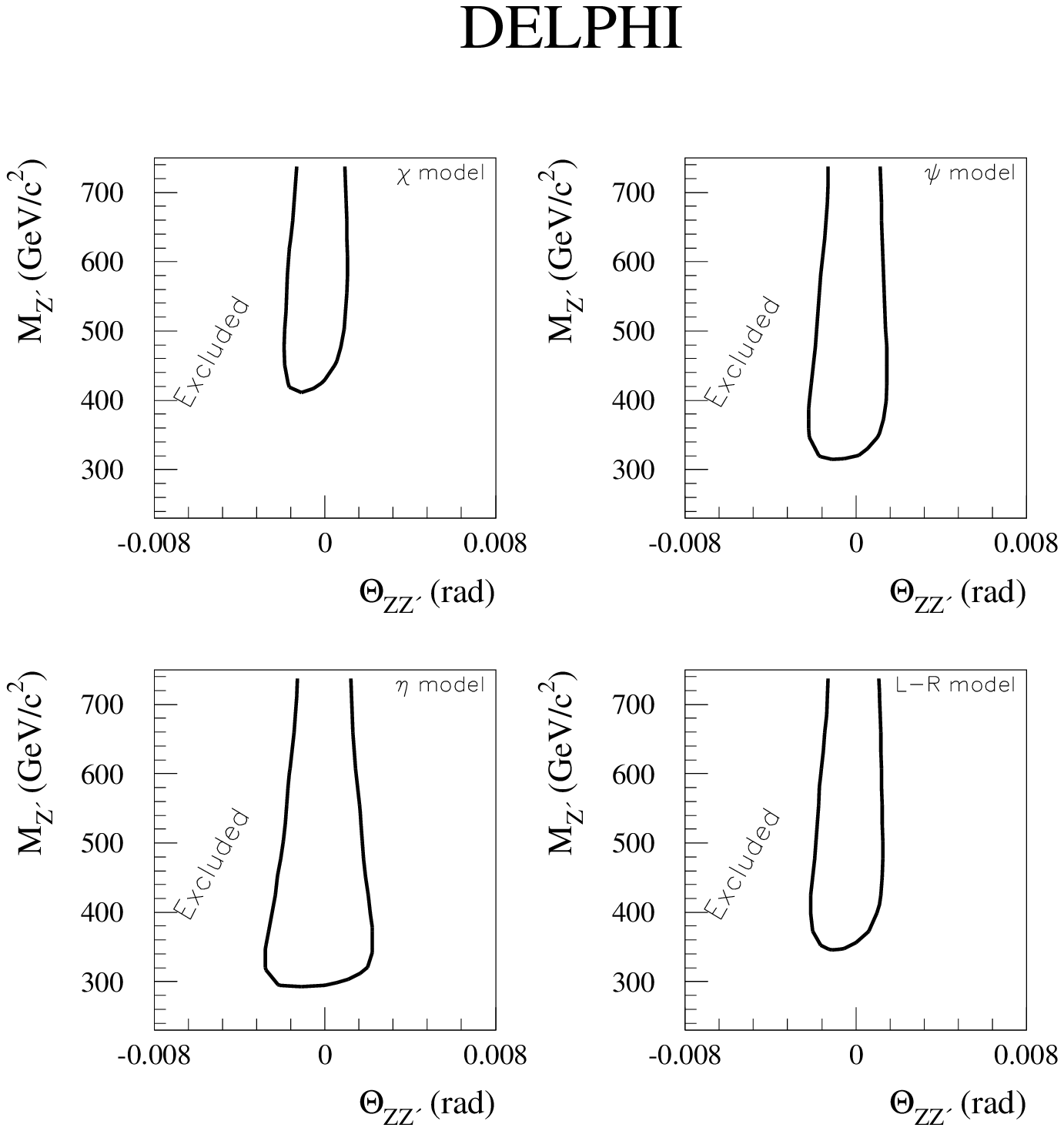,width=\textwidth}}   
  \caption[]{\capsty 
           {The allowed domain in the $\MZp - \thtzzp$ plane
           for the $\chi$, $\psi$, $\eta$ and L-R 
           models~\cite{ref:zprime-thry}. 
           The contours show the  95\% confidence level limits.}}
  \label{moddep_Mmix_limits}
 \end{center}
\end{figure}

Fits were made to the mass of a $\Zprime$, $\MZp$, the mass of the Z, $\MZ$, 
and to the mixing angle between the two bosonic fields, $\thtzzp$, for 
4 different models referred to as $\chi$, $\psi$, $\eta$ and 
L-R~\cite{ref:zprime-thry}. 
The theoretical prediction made came from the ZEFIT 
package~\cite{ZEFIT}. 
The fitted value of $\MZ$
was found to be in agreement with the value found from fits to the data
with no additional $\Zprime$.
No evidence was found for the existence of a $\Zprime$--boson 
in any of the models. 
The $95\%$ confidence level limits on $\MZp$, and $\thtzzp$,
were computed for the different model by determining the 
contours of the domain in the 
$\MZp - \thtzzp$ plane where $\chi^2 < \chi^2_{min} + 5.99$~\cite{ref:minuit}.
The allowed regions for $\MZp$ and $\thtzzp$ are shown in 
Figure~\ref{moddep_Mmix_limits}. The lower limits, shown in 
Table~\ref{tab_masslimits_moddep}, on the 
${\Zprime}$ mass range from 310 to 440 $\GeV/c^{2}$, an increase of 
between 70 and 190 $\GeV/c^{2}$ on the limits 
presented in~\cite{ref:delphi_130_172}, depending on the model.

In addition to the models considered in \cite{ref:delphi_130_172} a limit has
been obtained on the mass and mixing of the $\Zprime$ in the 
Sequential Standard Model \cite{ref:sqsm}. 
This model proposes the existence of a $\Zprime$ 
with exactly the same coupling to fermions as the standard Z.
A limit of \mbox{$\MZp > 710 \: \GeV/c^{2}$} is found at $95\%$ 
confidence level.

\begin{table}[p]
\begin{center}
\renewcommand{\arraystretch}{1.5}
\begin{tabular}{|cc|c|c|c|c|}
\hline
 \multicolumn{2}{|c|}{Model}          & $\chi$  & $\psi$ & $\eta$ & L-R    \\
\hline \hline
 $\MZplim$           & ($\GeV/c^{2}$) & 440     & 350    & 310    & 380    \\
\hline
$| \ \thtzzplim \ |$ & (radians)      & 0.0017  & 0.0018 & 0.0024 & 0.0018 \\ 
\hline
\end{tabular}
\end{center}
\caption{\capsty 
           {95\% confidence level lower limits on the $\Zprime$ mass and 
           upper limits on the Z$\Zprime$ mixing angle within the $\chi$, 
           $\psi$, $\eta$ and L-R models~\cite{ref:zprime-thry}. 
                }} 
  \label{tab_masslimits_moddep}
\end{table}

\subsubsection{Model independent fits}
%
%
%

Model Independent fits were performed to the leptonic cross--sections and 
forward--backward asymmetries, for the leptonic couplings of a $\Zprime$,
$a^{N}_{l'}$ and $v^{N}_{l'}$, normalised for the overall coupling scale 
and the mass of the $\Zprime$~\cite{ref:zp-modind-thry}. 

Several values of the mass of the ${\Zprime}$ 
were considered (i.e. 300, 500 and 1000 $\GeV/c^{2}$), and the 
Z${\Zprime}$--mixing was neglected. 
The limits on the normalised couplings are $|a^{N}_{l'}| < 0.15$ and 
$|v^{N}_{l'}| < 0.22$, a decrease of $0.04$ and $0.22$, respectively, 
on limits given in \cite{ref:delphi_130_172}.

\subsection{Gravity in Extra Dimensions}

The large difference between the electroweak scale 
($\mathrm{M_{EW}} \sim 10^{2} - 10^{3}$ GeV)  and the scale at which quantum 
gravitational effects become strong, the Planck scale
($\mathrm{M_{Pl}} \sim 10^{19}$GeV), leads to the well known 
``hierarchy problem''. A 
solution, not relying on supersymmetry or technicolour, has been 
proposed~\cite{ref:gravadd} that involves an effective 
Planck scale, $\mathrm{M_{D}}$, of 
$\mathcal{O}$(TeV). This is achieved by introducing $n$ 
compactified dimensions, into which spin 2 gravitons propagate, in addition to
the 4 dimensions of standard space-time. The Planck mass seen in the 4 
uncompactified dimensions, $\mathrm{M_{Pl}}$, can be expressed 
in terms of $\mathrm{M_{D}}$, the effective Planck scale in 
the $n+4$ dimensional theory,
\begin{displaymath}
\mathrm{M_{Pl}}^{2} ~ \sim ~ \mathrm{R}^n \mathrm{M_{D}}^{n+2} 
\end{displaymath}
where $\mathrm{R}$ is the size of the extra dimensions.
With $\mathrm{M_{D}}=1$~TeV, the case where $n=1$ is 
excluded as Newtonian gravitation would be modified at solar system distances 
whereas, $n=2$ corresponds to a radius for extra dimensions of 
$\mathcal{O}$(1 mm), which is not excluded by existing 
gravitational experiments \cite{ref:gravsubmil}.

In high energy collisions at LEP and other colliders, new channels
not present in the Standard Model would be available in which gravitons 
could be produced or exchanged.
Virtual graviton exchange would affect the differential cross section for
$\eeff$, with the largest contributions seen at low angles with respect to the 
incoming electron or positron. 
Embedding the model into a string model, and identifying the effective
Planck scale, $\mathrm{M_{D}}$, with the string scale, $\mathrm{M_{s}}$,
the differential cross section for $\eeff$ with the inclusion of the spin 2 
graviton can be expressed as \cite{ref:gravhew}:
\begin{displaymath}
\frac{\mathrm{d} \sigma}{\mathrm{d} \cos \theta}~
            =~A(\cos \theta)~
               +~B(\cos \theta)\left[\frac{\lambda}{\mathrm{M^4_s}}\right]~
                 +~C(\cos \theta)\left[\frac{\lambda}{\mathrm{M^4_s}}\right]^2,
\end{displaymath}
with $\theta$ being the polar angle of the outgoing fermion with respect to 
the direction of the incoming electron. The functions $A, B$ and $C$ are 
known, and the maximum power in the expansion is $\cos \theta^{4}$.
The dimensionless parameter $\lambda$, of $\mathcal{O}$(1), 
is not explicitly calculable
without full knowledge of the underlying quantum gravitational theory. It can 
be either positive or negative \cite{ref:gravhew,ref:gravgiu}. For 
the purposes of the fits, two cases, $\lambda = \pm 1$, are considered.
This parameterisation has no explicit dependence on the number of 
extra dimensions, $n$.

\begin{table}[p]
\begin{center}
\begin{tabular}{|c|c|c|c|} 
\hline
            & $\epsilon_{fit}$ &           & $\mathrm{M_{s}}$(TeV) \\ 
Final State & ( TeV$^{-4}$)    & $\lambda$ & $[95\% {\mathrm{C.L.}}]$ \\
\hline
\hline
\mumu
            & $-6.53^{+4.61}_{-2.24}$  &
                        $\begin{array}{c} -1 \\ +1 \end{array}$ & 
                         $\begin{array}{c} 0.559  \\ 0.649 \end{array}$  \\ 
\hline
\tautau
            & $-10.91^{+3.84}_{-8.18}$ &       
                        $\begin{array}{c} -1 \\ +1 \end{array}$ & 
                         $\begin{array}{c} 0.450  \\  0.564 \end{array}$ \\
\hline
\lplm
            & $-8.39^{+3.75}_{-1.96}$ &
                        $\begin{array}{c} -1 \\ +1 \end{array}$ & 
                         $\begin{array}{c} 0.542 \\ 0.680 \end{array}$   \\ 
\hline
\end{tabular}

\end{center}
\caption{\capsty{$95\%$ confidence level lower limits on $\mathrm{M_{s}}$ 
         in models of gravity in extra dimensions for $\mumu$ and 
         $\tautau$ final states, and for $l^{+}l^{-}$, a combination of both 
         muon and tau final states.}}
\label{tab:gravres}
\end{table}

\begin{figure}[p]
\begin{center}
\begin{tabular}{cc}
 \mbox{\epsfig{file=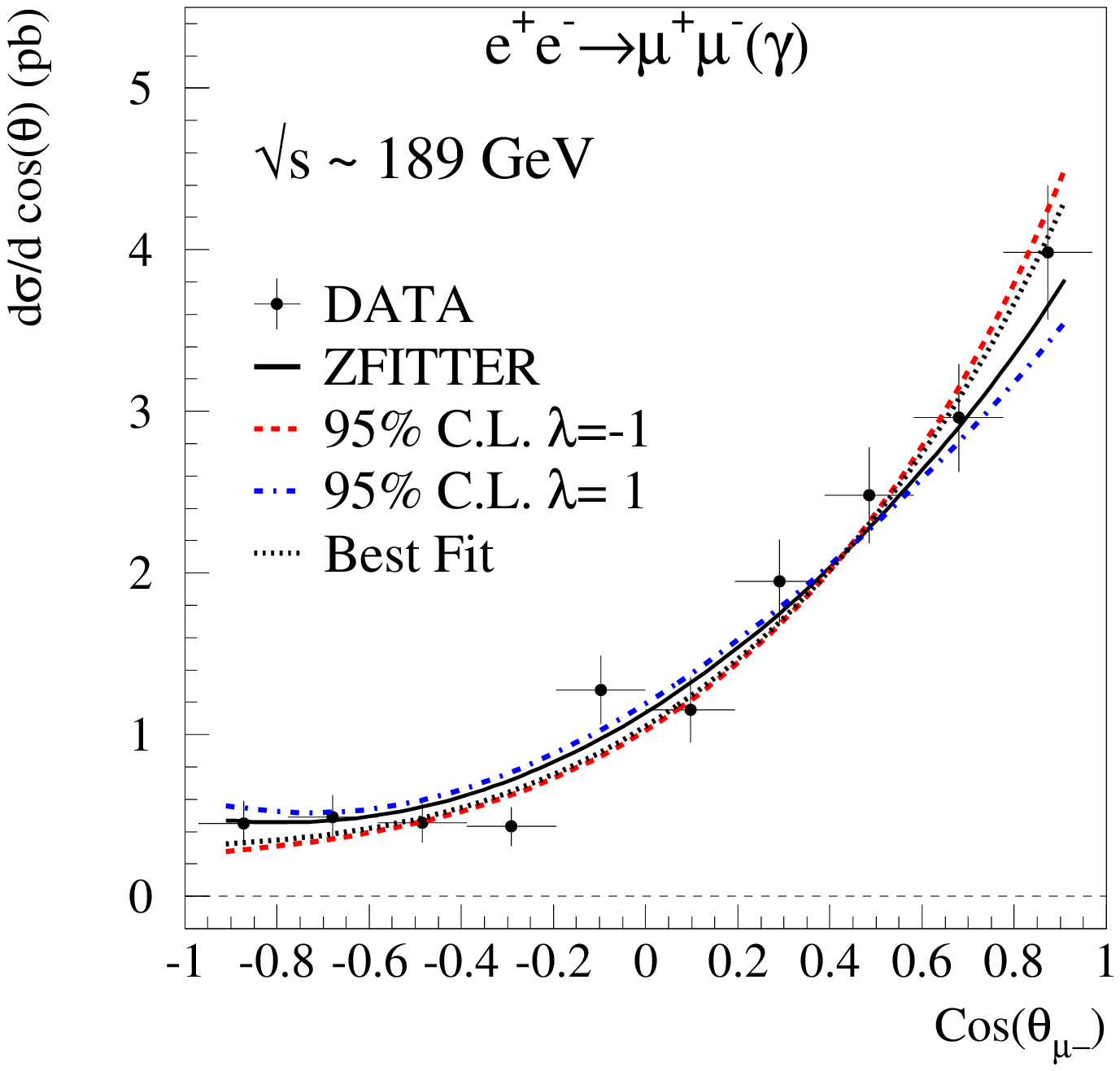,width=0.48\textwidth}} &
 \mbox{\epsfig{file=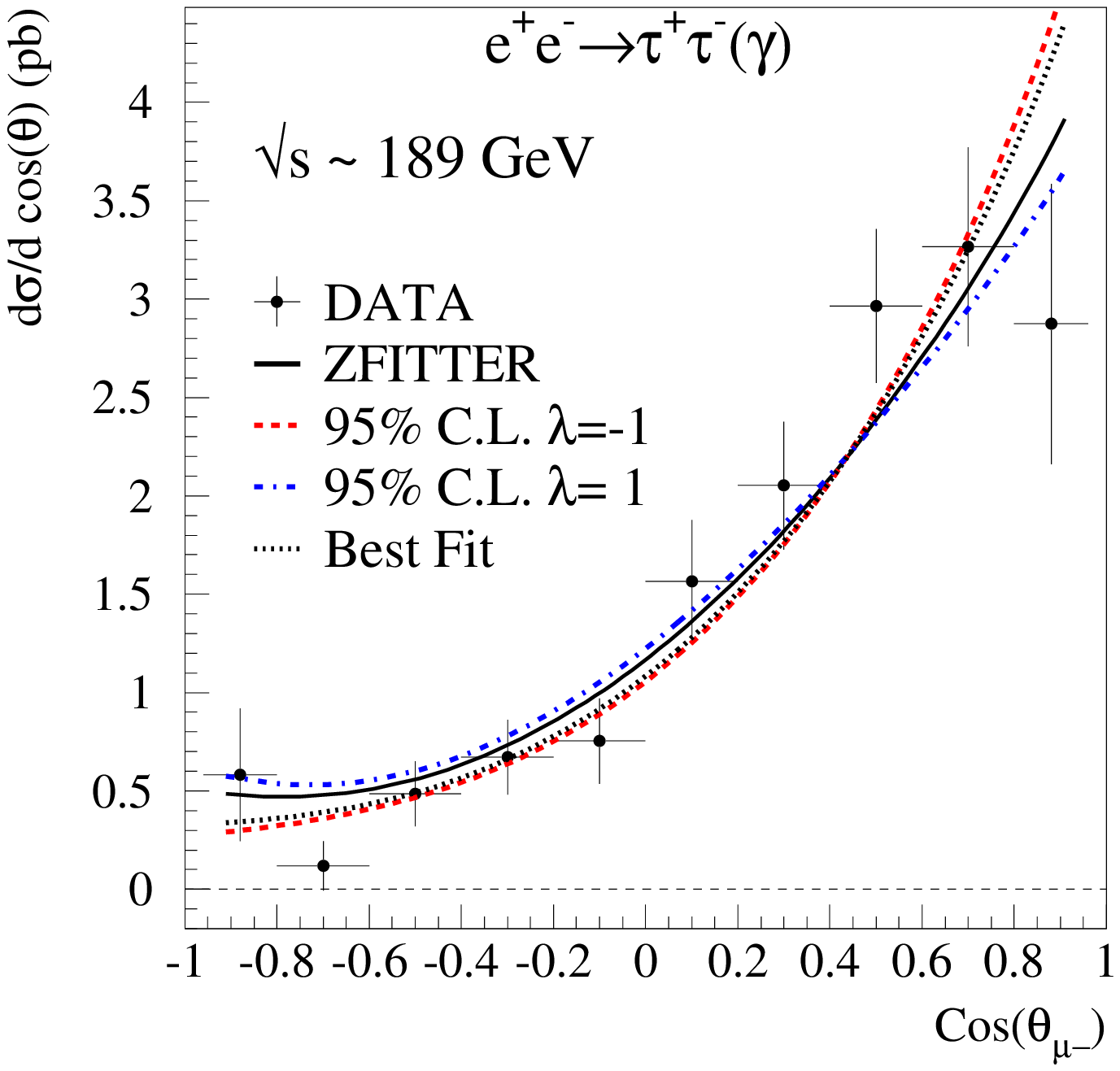,width=0.48\textwidth}} \\
\end{tabular}
\caption[]{\capsty{Fits to angular distributions for $\mumu$ and $\tautau$
final states, for models which include gravity in extra dimensions.
The dashed fitted curves correspond to $\epsilon = -8.39$~TeV$^{-4}$,
the {\emph{best fit}} to all data. 
The data are compared to the Standard Model predictions of ZFITTER 
and to the differential cross--sections predicted at $95\%$ 
C.L. for $\lambda=\pm 1$.}}
\label{fig:gravfit}
\end{center}
\end{figure}

Fits to the differential cross--sections, $\dsdcth$, measured at 
$\sqs \sim 183$ and 189 GeV
for the parameter $ \epsilon = \lambda / \mathrm{M^4_s} $ 
were performed, giving values compatible with the Standard Model, 
i.e. $\epsilon=0$. The systematics errors known to be fully correlated
between bins of $\cos\theta$ where treated as such.
Table~\ref{tab:gravres} shows the fitted values of $\epsilon$ and
$95\%$ confidence level lower limits on $\mathrm{M_{s}}$. These limits
were obtained using a method equivalent to that used to
extract the limits on the scale, $\Lambda$, of contact interactions, 
as described in section 6.1.1 of~\cite{ref:delphi_130_172}. 

The angular distributions predicted at $\sqs \sim 189 \GeV$
for the fitted values of $\epsilon$ are shown in Figure~\ref{fig:gravfit}.
The predictions for the values of $\mathrm{M_{s}}$ at the limits with 
$\lambda = \pm 1$, the data and the Standard Model predictions
are superimposed.

\section{Summary and conclusions}
\label{sec:summary}

The results of the analyses of cross--sections and asymmetries in the channels
$\eeeeg$, $\eemumug$, $\eettg$ and inclusive $\eeqqg$, 
at $\sqs \sim 183 - 189 \: \GeV$ have been presented. 
Overall, the data agree with the Standard Model predictions as calculated with
$\ZFITTER$ and $\TOPAZZERO$. 
The data were used to update previous searches for physics beyond the Standard
Model given and to investigate the possible effects of gravity in extra 
dimensions. No evidence for physics beyond the Standard Model 
was found and limits were set on parameters of several more general models. 
The scale $\Lambda$ 
characterising contact interactions between leptons can be excluded at $95\%$
confidence level in the range $\Lambda < 4.4 - 10.7 \; \TeV$ depending on the
model. For sneutrino exchange in R-parity violating supersymmetry, the 
generic coupling in the purely leptonic part of the superpotential,
$\lambda > 0.1$ can be excluded for $\msneut$ in the range 
130 - 190 $\GeV$ for all leptonic states at the $95\%$ confidence level or 
above. Alternatively, $\Zprime$ bosons lighter than $\sim 300 \: \GeV/c^{2}$ 
can be excluded at the $95\%$ confidence level in the models considered.
Lastly, $95\%$ confidence level lower limits of $542$ and $680$ GeV on 
the string scale, $\mathrm{M_{s}}$, in models of gravity involving 
extra dimensions are obtained for a combinations of 
$\mumu$ and $\tautau$ final states.

\subsection*{Acknowledgements}
\vskip 3 mm
 We are greatly indebted to our technical 
collaborators, to the members of the CERN-SL Division for the excellent 
performance of the LEP collider, and to the funding agencies for their
support in building and operating the DELPHI detector.\\
We acknowledge in particular the support of \\
Austrian Federal Ministry of Science and Traffics, GZ 616.364/2-III/2a/98, \\
FNRS--FWO, Belgium,  \\
FINEP, CNPq, CAPES, FUJB and FAPERJ, Brazil, \\
Czech Ministry of Industry and Trade, GA CR 202/96/0450 and GA AVCR A1010521,\\
Danish Natural Research Council, \\
Commission of the European Communities (DG XII), \\
Direction des Sciences de la Mati$\grave{\mbox{\rm e}}$re, CEA, France, \\
Bundesministerium f$\ddot{\mbox{\rm u}}$r Bildung, Wissenschaft, Forschung 
und Technologie, Germany,\\
General Secretariat for Research and Technology, Greece, \\
National Science Foundation (NWO) and Foundation for Research on Matter (FOM),
The Netherlands, \\
Norwegian Research Council,  \\
State Committee for Scientific Research, Poland, 2P03B06015, 2P03B1116 and
SPUB/P03/178/98, \\
JNICT--Junta Nacional de Investiga\c{c}\~{a}o Cient\'{\i}fica 
e Tecnol$\acute{\mbox{\rm o}}$gica, Portugal, \\
Vedecka grantova agentura MS SR, Slovakia, Nr. 95/5195/134, \\
Ministry of Science and Technology of the Republic of Slovenia, \\
CICYT, Spain, AEN96--1661 and AEN96-1681,  \\
The Swedish Natural Science Research Council,      \\
Particle Physics and Astronomy Research Council, UK, \\
Department of Energy, USA, DE--FG02--94ER40817. \\


\end{document}